\documentclass[12pt]{article}

\usepackage[T1]{fontenc}
\usepackage[scaled]{helvet}

\usepackage{hyperref}
\usepackage[headheight = 11ex,  top = 3.2cm]{geometry}

\usepackage{tgbonum}

\usepackage{titlesec}
\titleformat{\section}
  {\normalfont\normalsize\bfseries}{\thesection.}{1em}{}
\titleformat{\subsection}
  {\normalfont\normalsize\itshape}{\thesubsection.}{1em}{}
\titleformat{\subsubsection}
  {\normalfont\normalsize\itshape}{\thesubsubsection.}{1em}{}

\usepackage{amsmath,bm}
\usepackage{graphicx,pdflscape}
\usepackage[toc,page]{appendix}
\usepackage{simpler-wick}
\usepackage{etaremune,url,mdframed}
\usepackage{color,subfigure}

\usepackage{float} 
\usepackage{hyperref}
\hypersetup{
    colorlinks,
    citecolor=red,
    filecolor=cyan,
    linkcolor=blue,
   urlcolor=magenta
 }

\usepackage{xparse}
\ExplSyntaxOn
\NewDocumentCommand{\mref}{m}{\quinn_mref:n {#1}}
\seq_new:N \l_quinn_mref_seq
\cs_new:Npn \quinn_mref:n #1
 {
  \seq_set_split:Nnn \l_quinn_mref_seq { , } { #1 }
  \seq_pop_right:NN \l_quinn_mref_seq \l_tmpa_tl
  ( 
  \seq_map_inline:Nn \l_quinn_mref_seq
    { \ref{##1},\nobreakspace } 
  \exp_args:NV \ref \l_tmpa_tl 
  ) 
 }
\ExplSyntaxOff

\newcommand{\disp}[1]{Eq.~\mref{#1}}

\newcommand{\figdisp}[1]{Fig.~\mref{#1}}

\newcommand{\lessim} {\ {\raise-.5ex\hbox{$\buildrel<\over\sim$}}\ }

\newcommand{\gssim}{\ {\raise-.5ex\hbox{$\buildrel>\over\sim$}}\ }

\newcommand{\si}{\sigma}
\newcommand{\sib}{\bar{\sigma}}
\newcommand{\tJ}{\ $t$-$J$ \ }
\newcommand{\nn}{\nonumber}

\newcommand{\wt}{\widetilde}

\renewcommand{\Im}{\mathrm{Im}}
\newcommand{\Tr}{\mathrm{Tr}\,}

\renewcommand{\emph}{\textit}

\newcommand{\lll}{\langle \langle}

\newcommand{\rrr}{\rangle \rangle}

\newcommand{\utoo}{\frac{u_0}{2}}

\usepackage{hyperref}
\usepackage{color}

\newcommand{\bitem}{\begin{itemize}}
\newcommand{\eitem}{\end{itemize}}
\newcommand{\bmu}{{\bm{ \mu}} }
\newcommand{\beq}{\begin{eqnarray}}
\newcommand{\eeq}{\end{eqnarray}}
\newcommand{\barray}{\begin{eqnarray}}
\newcommand{\earray}{\end{eqnarray}}

\newcommand{\ang}{\text{\AA}}

\usepackage{authblk}
\graphicspath{ {Figures-Overview/} ,{../Figures-Overview/}}

\usepackage{changepage}

\newcommand*{\bigchi}{\mbox{\large$\chi$}}
\newcommand{\htsc}{High-T$_c$ }

\begin{document}
\thispagestyle{empty}


\begin{center}
{\bf  \Large Overview of the \\Theory of Extremely Correlated Fermi Liquids }
\vspace{.5 in}\\
B Sriram Shastry\footnote{sriram@physics.ucsc.edu}\\
University of California Santa Cruz\\
Santa Cruz, CA 95064\\
{\it 5 November 2026}
\end{center} 

\vspace{1in}

\begin{center}
{\bf  Abstract}
\end{center}

The Extremely Correlated Fermi Liquids (ECFL) theory is reviewed as a framework for understanding the \(t\)-\(J\) model in metallic systems close to the Mott insulating limit. This  overview presents the  underlying ideas  and the resulting equations in 
a form accessible to nonexperts. We compare theoretical results with all available resistivity data for single-layer High-\(T_{c}\) systems, and with some spectral data.  The    highlighted results include  a  density dependent quasilinear \(T\)-dependence in resistivity, an unusually  small quasiparticle weight, and distinct low-temperature emergent scales that dominate transport, thermodynamics  and spectral properties of  single-layer High T$_c$ systems. Suggestions are made for further experiments to probe the physics of these challenging quantum 
many-body  systems.
\newpage
\pagestyle{plain}

\section{Introduction\label{Intro}}
In this article I present an overview of  the theory of  Extremely Correlated Fermi  Liquids  (ECFL) \cite{ECFL} introduced by the author in 2011. This theory has been developed in order to provide a systematic understanding of the physics of very strongly interacting electrons in  a narrow band. Such a  setting occurs in nature in the important class of strongly correlated metallic systems, which  includes  High T$_c$ superconducting materials.  It defines a notoriously difficult  regime, where the  assumption of  the text-book  Feynman-Dyson perturbation theory, of a small interaction constant, is no longer true. The results based on this approach are therefore of dubious value. Inventing and developing  alternate non-perturbative approaches  is  essential to make headway in this class of problems. This  is widely recognized as  a central problem in condensed matter physics. The ECFL theory is precisely  such a theory; in this theory a systematic method is developed, which has the potential to give increasingly accurate results as one proceeds  along  the route developed in the scheme.    The \tJ model  is the immediate goal for this approach. This model  is closely related to the Hubbard model in the limit of almost  infinitely strong interactions.  Their mutual relationship  has been well  understood for decades \cite{tJ-Hubbard-Connection},  but neither model is easy to solve in the relevant regime of parameters. 

Since the  inception of the ECFL theory in 2011, a considerable body of results has emerged \cite{ECFLall}.  The goal of this article is to provide    a broad picture of the   goals of the theory, to  summarize its formulation and to display  some highlights  of results obtained so far. This article is designed to be a ready reference to this body of work, and to  provide pointers to papers containing further details. Some effort has been  made to make this article accessible to  advanced graduate students and postdoctoral fellows, by providing  extended discussions when possible.  This   article  is { not} designed  to be a  review of  the vast field of strongly correlated electron systems, which has  seen the introduction of several novel  ideas in theory and refinements in experimental techniques in recent years. These  recent  developments are too varied for a brief and balanced summary. Instead,  in this article I review the body of work done by our group \cite{ECFLall}  emphasizing the basic  underlying  theme, and  showcase some  interesting results.  The reader will find  pointers to several benchmarking  results  in Section~\ref{others}, where the ECFL theory has been tested against other  well established  techniques  having overlapping results.  The article also contains detailed comparisons of our  results with a range of published experimental data  using various probes, without any claim to being a thorough review of data either.

The setting to the approach advocated in the ECFL theory is the explosive interest in strongly correlated systems after the discovery of \htsc phenomenon in 1986.  The  discovery of superconductivity in the cuprate systems  is remarkable  due to its proximity to a Mott insulating state- a hallmark of strong correlations. This was soon followed by two other notable results, namely the discovery of quasi-linear resistivity $\rho(T)$$\sim$$T$ observed \cite{Martin} in the  Bi2201 system (Bi$_2$Sr$_2$CuO$_{6+x}$),  and the apparent absence of quasiparticle peaks coupled with  extremely  wide spectral functions seen \cite{Arko} in angle resolved photoemission studies (ARPES) in the Bi2212 system (Bi$_2$Sr$_2$CaCu$_2$O$_{8+x}$ ). 

Both of these results seemed to indicate a possible breakdown of the foundational  Landau's Fermi liquid theory of metallic systems, where one expects a different behaviour with $\rho(T)$$\sim$$T^2$. In ARPES  studies of Fermi liquids, one expects to see very sharp quasiparticle peaks in the spectral function, with a compact background not extending beyond the  bandwidth (typical $\sim$2eV in these systems). 

In the context of resistivity, Fermi liquids with significant electron-electron interactions  are usually expected to display a prominent  $\rho(T)$$\sim$$T^2$ behaviour.  This  was  already predicted by Landau and Pomeranchuk \cite{Landau-Pomeranchuk} in  1937. They noted that electron-electron scattering would  lead to a $T^2$ contribution to  the resistivity of  metals, and on adding the phonon scattering contribution, would give a $\rho\sim \alpha T^2+ \beta T^5$ behavior at low temperatures. The key point made in \cite{Landau-Pomeranchuk}, which is worth recalling,  is that in a  metal with  electrons moving in a Bloch band,  the conservation of the total  crystal momentum of a pair of electrons in the scattering process does not imply the conservation of their total {group velocity}, which enters the transport equations.  This allows a certain fraction of the scattering processes-
 the  {\em umklapp} processes- with a non-zero reciprocal lattice vector,  to  balance the momentum,  while allowing a non-zero transfer of velocity, leading to a non-vanishing resistivity. The fraction of umklapp processes is almost unity for narrow band strongly correlated systems \cite{Wilkins,MJRice,Miyake-Varma}, and therefore this process is very efficient.
 
 Thus the observed departure from the $\rho(T)$$\sim$$T^2$  behaviour could be taken  to indicate a breakdown of the Fermi liquid theory, and led to  much activity in the community.
 
 To summarize, the above early  data  posed  two main questions
 \begin{itemize}
\item[ (Q1:)] What is the physics leading to the unusual T dependence of the resistivity and the broad and featureless spectral lines seen in ARPES studies?
 \item[(Q2:)] Is the mechanism for the observed d-wave  superconductivity related to the  physics leading to the   unusual resistivity and photoemission?
\end{itemize}

The ECFL theory was  formulated in response to the above challenge. The technical roadblock in formulating  a theory for 
 strongly correlated metals is absence of any small parameter in the model. The usual method in many-body systems  of using the interaction  strength as a perturbative parameter is clearly undermined, if the interaction strength  is very large, even bigger than the band width. Under these conditions  perturbative methods, including those that resort to summation of selected set of infinite  diagrams, become  dubious. This has created a large theoretical gap in the space of techniques, wherein the ECFL theory has been launched. 
 
  The ECFL theory  aims at providing:
\begin{itemize}
\item[(a)] An  analytical treatment of t-J  model. 

\item[(b)]  Systematic approximations starting from the Fermi gas limit.

\item[(c)] Systematic calculations of  the one electron spectral function $A(k,\omega)$, the  resistivity, Hall response etc.  as functions of T, and particle density n (the number of electrons per lattice site)\footnote{ In this paper I will refer to the electron density as n, and the hole density as $\delta$, related  to n by $\delta$=1-n. 
}. 
   \item[(d)] Studying the possibility of a superconducting instability and  calculating the transition temperature Tc. 
 \end{itemize}
 Towards goal (a) the ECFL theory provides approximations to increasing orders in a parameter $\lambda$, which is noted to be analogous to the inverse spin $\frac{1}{ S}$ in a semi-classical approximation of  interacting quantum spins.  In goal (b) the lowest order theory is just the free Fermi gas, and terms of higher order in $\lambda$ build in correlations continuously, while preserving the volume of the Fermi surface. This implies that the successive approximations satisfy the Luttinger-Ward theorem\cite{Luttinger-Ward}. We take this feature to be  an essential requirement for a satisfactory theory for cuprate systems, keeping in view  their observed compliance with this theorem.  In goal (c), the  parameters of the t-J model in 2-d are chosen from  the actual description of High T$_c$ cuprates with a single Cu-O layer per unit cell. Different sets of parameters are obtained for various systems from corresponding  experiments. This allows the theory to make specific  predictions for each  system, which can  then be tested against data. Goal (d) is   more open-ended  than the others. Preliminary results in this direction are noted here, while further work is underway.

\section{The Hubbard and t-J models}
 
 The most popular model for incorporating strong correlation effects is undoubtedly the Hubbard model. Here one simply adds a correlation term $U \sum_i n_{i \uparrow} n_{i \downarrow}$ to the band structure determined kinetic energy term $T= \sum_{k \si} \varepsilon_k C^\dagger_{k \si} C_{k \si}$.The Hubbard model provides a very effective description of  weak or intermediate coupling situations where $U\lessim W$ ($W$ is the band width).  
 The real difficulty of dealing with  this model arises  for strong correlations, which may be loosely   defined  as  e.g.  $U\geq W$.
 Near the insulating limit of $\delta$$\sim0$, this qualitative criterion becomes  $U\geq \delta \; W$, making things worse as far as theory is concerned, since
 in this regime perturbation theory is unmanageable. 
 
 The special  case of 1-d has the well known exact solution using the Bethe Ansatz,  for nearest neighbour hopping. The solution  is valid for any value of $U/W$, and provides a classic example of the Mott insulator.  In the case of large d, one  has a satisfactory  numerical  solution for any value of $U/W$ in terms of the dynamical mean field theory (DMFT) \cite{DMFT}. This theory  is based on the $\vec{k}$  independence of the Dyson self energy, and  leverages the solution of the single site Anderson impurity model available from Wilson's renormalization group   solution of that problem \cite{NRG,NRG2}. For d=2, which is of  particular interest for understanding cuprate High T$_c$ materials, analytical theories in the strong correlations regime  are hard to find.

Another popular problem is the \tJ model,  which has features that  distinguish it from  the Hubbard model.
 Its great advantage over the Hubbard model arises from the fact that the  large energy scale $U$ is removed at the very outset using a canonical transformation. In this scheme, the large size of  U/W leads to  the Gutzwiller  constraint of no-double occupation, and an added  superexchange antiferromagnetic  term $J \sum \vec{S}_i.\vec{S}_j$ \cite{tJ-Hubbard-Connection}, with  $J= \frac{t^2}{4 U}$. The residual degrees of freedom are the delocalized Gutzwiller projected  electrons, with non-canonical anti-commutation relations, which give rise to the Fermi surface and transport properties. In summary an electron of the weakly interacting problem gets  (dynamically ) decomposed into two pieces,  a longlived spin moment plus  a mobile charge. The mobile charge in turn is decomposed into a quasiparticle part and an incoherent background part. A theory for the \tJ model is required to describe this picture in a quantitative fashion.
In the ECFL theory we choose to work with the \tJ model \disp{HtJ}, with a systematic scheme for calculating physical quantities were such a decomposition is implicitly achieved.

\subsection{ \tJ Model and the  Luttinger-Ward Fermi surface theorem.}

The \tJ model may be regarded from one of two viewpoints- as a standalone model without any small parameter, or as a limiting case with $U\to\infty$ of the Hubbard model. In the first viewpoint  we have relatively few general constraints on the solution at T=0- even if we assume that  it can solved. In viewing it as a limiting case of the Hubbard model it is natural   to extend  to the solution of the \tJ model some general  features originally found in  weak coupling analysis, most notably the important T=0 Fermi surface volume theorem of Luttinger and Ward (L-W) \cite{fn1,Luttinger-Ward,Shastry-Luttinger,Oshikawa-Luttinger}

 Implicit in the above discussion is the idea that the \tJ model could in principle have  several  classes of   solutions  amongst which  one class would  satisfy the L-W theorem and another may not satisfy the L-W theorem as T$\to$0. Given such a choice one might favour selecting the solution giving the lowest free energy. However 
  picking between these different  classes of solutions on the basis of minimizing the free energy is not possible, since  this object cannot be calculated reliably. The situation is further complicated by the possibility that  small added terms in the Hamiltonian could  lead to  different answers.   This leaves us with a fork in the road, with significant differences in the  development of a suitable theory.  I adopt the  first path by requiring  that satisfying the L-W  Fermi volume theorem is a necessary ingredient  for the  ECFL theory. This choice is guided by the physics of the problem,  since our   goals require  paying  close attention to the observed, Luttinger-Ward compliant Fermi surfaces of several High $T_c$ cuprates.  For some cuprates, notably in the underdoped regimes, the Fermi surface  is reported  to break up into disconnected segments, and may be interpreted as violating the  Luttinger-Ward theorem. We do not currently have a theory valid  for the underdoped case with $\delta$$\ll$1, which is required in such  cases, and therefore shall  not attempt to explain them.

It should be  mentioned that considerable work on the \tJ model has taken the other path, one giving a different volume of the Fermi surface \cite{Izumov,ECQL}. These treatments amount  to a re-summation of an expansion in powers of $\beta t$ and  $\beta J$. Since the expansion parameter is proportional to $\beta$, these can be argued to correspond to a high temperature expansion, about the  highly degenerate (and highly   symmetric) $T=\infty$ state, where every (Gutzwiller constraint satisfying) state is allowed.  From this starting point,   obtaining  a Luttinger-Ward compliant  Fermi surface  at $T=0$  seems like a very difficult- or even impossible  task. The situation is somewhat  analogous to arriving at   a broken symmetric state in an antiferromagnet  from the high T state, without the help of a sufficiently small symmetry breaking term;  this  is  not a natural possibility. As an example of this second route,  by  using an entirely different scheme (termed the  ECQL scheme) for the \tJ model in d=2  in Ref. \cite{ECQL},  the author  found a large Fermi surface,   
occupying  an area $A_{\scriptscriptstyle FS}^{\scriptscriptstyle ECQL}= \frac{n}{2-n}$, where $n$ is the areal density. This is always larger than  the Fermi gas area $A_{\scriptscriptstyle FS}^{\scriptscriptstyle LW}=\frac{n}{2}$.  In the ECFL theory we ensure that we  get $A_{ \scriptscriptstyle FS}^{\scriptscriptstyle ECFL}=\frac{n}{2}$ at $T=0$.  As discussed here this is achieved by  introducing a continuous   parameter $\lambda\in[0,1]$ connecting the Fermi gas to the  \tJ model, and organizing an expansion in powers of $\lambda$ of the relevant self energies.

 \subsection{Formulating the ECFL  solution of the \tJ model }
Let us recapitulate the definition of the \tJ model. The Hamiltonian in the grand canonical ensemble  is written in terms of Gutzwiller projected Fermions $\hat{C}$ defined by
\beq
&&\widetilde{C}_{i \si}= P_G C_{i \si} P_G, \mbox{   and    } \widetilde{C}^\dagger_{i \si}= P_G C^\dagger_{i \si} P_G, \mbox{  with  } \nn \\
&&P_G= \prod_i(1-n_{i \uparrow} n_{i \downarrow}),
\eeq
as
\beq
H_{tJ}= - \sum_{ij} (t_{ij}) \wt{C}^\dagger_{i \si} \wt{C}_{j \si} -  {\bm{ \mu}} \sum_{i \si} n_{i \si} + J \sum_{<ij>}\left( \vec{S}_i.\vec{S}_j - \frac{1}{4} \rho_i \rho_j\right) \label{HtJ}
\eeq
where $n_{i \si}=\wt{C}^\dagger_{i\si}\wt{C}_{i\si}$  is the number operator, $\bmu$ the chemical potential, $\vec{S}_j$ the electron spins and $\rho_i=\sum_\si n_{i \si}$, and the summation symbol $<ij>$ denotes a sum over each nearest neighbour bond.
The kinetic energy term is defined with arbitrary hopping parameters $t_{ij}$, and upon Fourier transforming becomes the familiar expression $\sum_{k \si} (\varepsilon_k- \bmu)C^\dagger_{k \si} C_{k \si}$ for an uncorrelated band, where $\varepsilon_k$ is the band energy.
 The $\wt{C}$ operators (sometimes known as the Hubbard $X$ operators) satisfy anti-commutation relations
\beq
\{ \wt{C}_{i\si_1}, \wt{C}^\dagger_{j \si_2} \}=\delta_{ij}(1- \si_1 \si_2 \wt{C}^\dagger_{j \sib_1} \wt{C}_{j \sib2} ) \label{ACR},
\eeq
where we denote reversed spins by the symbol $\sib_j = -\si_j$. We note that due to the Gutzwiller constraint, the kinetic energy term is  by no means a simply solvable problem, unlike in the Hubbard model, and the exchange (i$J$)  term is also not  small. This is therefore an intrinsically strong coupling problem.

The ECFL solution to this problem\cite{ECFL,ECFLall} proceeds in four distinct steps which are as follows
\begin{itemize}
\item[Step(I)] {\em Find exact equations of motion for the single electron Greens function, and identify suitable self energies.} This is most easily done with the help of a Grassmanian path integral formulation and generates {\em two } self energies. This method \cite{PathIntegral} is simpler than the Schwinger method of sources used in \cite{ECFL}

\item[Step(II)] {\em Introduce a Lagrange multiplier $u_0$ which imposes the ``shift-invariance'' on the theory at any order in $\lambda$.} This shift invariance is the statement that  moving the center of gravity of the Fermi bands, i.e. adding a $\vec{k}$-independent constant to the band dispersion, should not affect the results of the calculation.

\item[Step(III)] {\em Introduce a parameter $\lambda$ in the path integral formulation and the Greens functions. }
The parameter is such that at $\lambda=0$  we find a free Fermi gas, while at $\lambda=1$ we obtain the \tJ model.  Expansions in powers of $\lambda$  of the self energies to a given order, give rise to systematically improvable  approximations.
 
\item[Step(IV)] {\em Determine the two self energies to any order in $\lambda$ from the equations.} The Greens function has a specific representation in terms of the two self energies mentioned  in the first step. 
\beq
G^{(\lambda)}_\si(\vec{k},\omega)= \frac{1- \lambda \frac{n}{2} + \Psi_\lambda(\vec{k},\omega)}{g_0^{-1}(\vec{k},\omega) -\Phi'_\lambda(\vec{k},\omega) },  \label{formofG}
\eeq
where $g_0$ is the Greens function for non-interacting Fermions. We are interested in the limit $\lambda$$=1$ for the \tJ model. At $\lambda=0$ we obtain the free electron limit, provided $\Psi_\lambda,\Phi'_\lambda$ vanish here. 
The Greens function can be considered as a product of an auxiliary Greens function $g$ given by  $g=\{ g_0^{-1} -\Phi'_\lambda   \}^{-1} $ with a Dyson type self energy $\Phi'$,  and $\wt{\mu}=\{1-\lambda n/2+ \Psi_\lambda\} $, determined by a second self energy $\Psi$. The second  term can be thought of as a caparison function, i.e. one  which provides a second layer of dressing beyond the one provided by the self-energy $\Phi'$.

\end{itemize}  
We address each of these steps next.

{\bf \S}{For implementing  Step(I)}
we   first describe a  path integral method for obtaining the exact equations of motion for the Greens function. This method is perhaps  preferable to   the methods  used in the original calculations \cite{ECFL}  for pedagogy, and leads to equivalent results.
The details can be found in \cite{PathIntegral}, so we will be brief.  Our first goal is to calculate the partition function 
\beq
{\cal Z}&=&\Tr P_G e^{- \beta H}\nn \\
&=& \lim_{M\to \infty} \Sigma <\nu_0| P_G e^{-\frac{\beta H}{M} }| \nu_1><\nu_1|  e^{-\frac{\beta H}{M} }|\nu_2>\ldots<\nu_{M-1}| e^{-\frac{\beta H}{M} }|\nu_0>\nn \\  \label{Part1}
\eeq
where subdivision of $e^{- \beta H}$ followed by the summation  over the $M$ complete sets of states $\{\nu_j\}$, is introduced to facilitate the use of Totter's formula and for taking the time continuum limit. Calculations are considerably simplified if we observe that the intermediate states can be  regular Fermionic states- without removal of doubly occupied states (henceforth   doublons), since (a) $H$ as given in \disp{HtJ}, does not create doublons when acting on states without doublons and (b)  $P_G$ which commutes with $H$ (and therefore can be placed either to the left or right of the thermal factor) filters out states with doublons present in the initial state $<\nu_0|$.

Let us consider   replacing  $\wt{C}^\dagger$ and ${\wt C}$ in \disp{HtJ} by much simpler operators 
\beq
\wt{C}_{i \si} \to C_{i \si},   \mbox{    and  }\wt{C}^\dagger_{i \si} \to C^\dagger_{i \si} \left( 1-n_{i \sib}\right)  \label{unhat}.
\eeq
It follows that with $H$ consisting of these replacements in \disp{HtJ}, and with a state $|\nu>$ that has no doublons (i.e. is Gutzwiller projected)  then the resulting  state  $e^{-\frac{1}{M} \beta H}|\nu>$ also  has no doublons. The reason is that the destruction operators $C_{i \si}$  removes a particle and can therefore not add a doublon to the state $|\nu>$, and the creation type operator in \disp{unhat} explicitly prevents adding a particle at a previously occupied site.     This crucial observation implies that {\bf all} the intermediate states in \disp{Part1} can be taken to be all  states  without bothering to eliminate the doublons. The matrix elements in the product structure  will  eliminate the states with doublons, since the product begins with  $P_G |\nu_0>$, a Gutzwiller projected state.

 We  may then use standard Fermionic coherent states and retain the projection operator at one end of the product. The partition function is now  a  functional  integral over all variables $c,c^*$
 \beq
 {\cal Z}= \int_{c,c^*} P_G(0) e^{-{\cal A}_{Tot} } \label{Part2}
 \eeq 
  where the action variable 
   \beq {\cal A}=  \int_0^\beta {\cal H}(c,c^*,\tau) d\tau, \label{Action1}
   \eeq is the sum of various terms
 \beq
 {\cal H}&=& {\cal H}_0+{\cal H}_t+{\cal H}_J \label{hambreak}
 \eeq
 with
 \beq
 {\cal H}_0=\sum_l  c^*_{l \si}(\tau) \left(\partial_\tau - { \bmu}  \right) c_{l \si}(\tau)
 \eeq
\beq
{\cal H}_t= - \sum_{lm\si}  t_{lm} c^*_{l \si}(\tau)c_{m \si}(\tau)\times (1- n_{l \sib}(\tau) ) \label{Ht}
\eeq
and 
\beq
{\cal H}_J= -\frac{1}{4} \sum_{lm \si} J_{lm} \si_1 \si_2 c^*_{l \si_1}(\tau)c^*_{m \sib_1}(\tau)c_{m \sib_2}(\tau) c_{l\si_2}(\tau). \label{HJ}
\eeq
The objects $c,c^*$ are Grassman variables corresponding to the eigenstates of $C$ and $C^\dagger$ in the coherent states, i.e. $C_{i \si} |\nu_j>= c_{i \si}(\tau_j) |\nu_j>$, and  $<\nu_j| C^\dagger_{i \si}=<\nu_j| c^*_{i\si}(\tau_j) $, and $\tau_j= \frac{j }{M} \beta $. 
 As $M\to \infty$ we take the the time continuum limit $\tau_j \to \tau \in [0,\beta]$. 
The products at equal times are required to be point-split- for example $c_a^*(\tau)c_b(\tau)$ represents  the product  $c_a^*(\tau_{j+1})c_b(\tau_j)$. Further details of this formulation are  detailed in \cite{PathIntegral} (Eq.~(108)).

{\bf\S} {For implementing Step(II)} we first  discuss the introduction of a useful   parameter  $u_0$ in the action, and the shift identity underlying it.  In the definition of the \tJ model \disp{HtJ}, a shift   transform
\beq
H &\to& H+ \utoo {\bm{Q}} \label{u01} \\
{\bm{Q}}&=&\sum_{i \si}\left(n_{i \si} -\sum_j \delta_{ij} \wt{C}_{i \si} ^\dagger \wt{C}_{j \si}  \right) \label{u02}
\eeq
with an arbitrary $u_0$ makes no difference if we treat the model exactly. This is evident since $\lim_{j\to i}  \wt{C}_{i \si} ^\dagger \wt{C}_{j \si} -n_{i \si}$, and hence $\bm{Q}=0$. However in  approximations made here, and indeed in similar calculations\cite{PauliPrinciple}, this limiting identity is not satisfied,  and   give a non-vanishing value for  the expectation value of $\bm{Q}$. For this purpose we propose to take \disp{u01} as the Hamiltonian,  where  a vanishing  average, i.e.  $\langle \bm{Q}\rangle=0$ is the constraint. Requiring it to hold  within a given approximation determines  the Lagrange multiplier  $u_0$, since the average is now a non-trivial  function of $u_0$.  This constraint  can be  added  by shifting  the hopping parameters as
\beq
t_{ij}\to t_{ij} +  \delta_{ij} \utoo,
\eeq
and simultaneously shifting the chemical potential $\bmu\to \bmu- \utoo $.

{\bf \S} For implementing Step (III), we   discuss the introduction  of a  parameter $\lambda$ in the action. The parameter $\lambda$  is introduced primarily in \disp{Ht} by writing $[1- n_{l \sib}(\tau)]\to [1- \lambda n_{l \sib}(\tau)]$. Setting $\lambda=0$ gives us the usual kinetic energy of  free  electrons, while $\lambda=1$ yields  the \tJ model. Since we want ${\cal H}$ to represent non-interacting  electrons
as $\lambda\to 1$, we also multiply \disp{HJ} by a factor $\lambda$

With these changes the partition function of interest is found from \disp{HtJ} with the changes
\beq
{\cal H}_0&\to&{\cal H}'_0= \sum_l  c^*_{l \si}(\tau) \left(\partial_\tau - { \bmu} +\utoo  \right) c_{l \si}(\tau) \nn \\
{\cal H}_t& \to& {\cal H}'_t= - \sum_{lm\si} ( t_{lm}+\utoo) c^*_{l \si}(\tau)c_{m \si}(\tau)\times (1- \lambda n_{l \sib}(\tau)) \label{Hkin}  \\
{\cal H}_J&\to&{\cal H}'_J=\lambda {\cal H}_J \label{HtJ2}
\eeq
so that ${\cal H}'={\cal H}'_0+{\cal H}'_t+{\cal H}'_J$.
Summarizing, we write the modified action 
\beq
{\cal A}'=  \int_0^\beta {\cal H}'(c,c^*,\tau) d\tau, \label{Action2}
\eeq

Turning to dynamics,  the one electron Greens function (including the Gutzwiller projection physics) is treated as follows
\beq
G_{i\si j \si'}(\tau,\tau')&=&- \frac{1}{{\cal Z}}\Tr T_\tau  \left( e^{-{\cal A'}} \wt{C}_{i \si}(\tau) \wt{C}^\dagger_{j \si'}(\tau') \right) \label{G1} \\
&=&\lll \left\{ 1- \lambda n_{j \sib'}(\tau') \right\} c^*_{j \si'}(\tau')c_{i \si}(\tau)\}\rrr \label{G2}
\eeq
where the second line represents the average calculated with the modified action  \disp{Action2}.

At this stage we can calculate the equations satisfied by $G$ using the Fermionic Grassman identity 
\beq
0=\int_{c c^*} {\cal P}_G \frac{\delta}{\delta c_{i \si}^*(\tau)}\left[ \left\{ 1- \lambda n_{j \sib'}(\tau')\right\} c^*_{j \si'}(\tau') e^{-{\cal A}'}\right]. \label{Fermidentity}
\eeq
This identity follows from the observation that for Grassman variables integration is the same thing as taking  a derivative, and taking two derivatives with respect to the same variable gives zero.  Since the integration implicit in \disp{Fermidentity} already provides one derivative, the second derivative gives a vanishing result. Dividing by ${\cal Z}$ this gives
\beq
\lll \frac{\delta}{\delta c_{i \si}^*(\tau)} [ \left\{ 1- \lambda n_{j \sib'}(\tau')\right\} c^*_{j \si'}(\tau')] \rrr=- \lll [ \left\{ 1- \lambda n_{j \sib'}(\tau')\right\}   c^*_{j \si'}(\tau') \frac{\delta}{\delta c_{i \si}^*(\tau)} {\cal A'} \rrr
\eeq
This gives the basic equation for G and can be simplified further as
 detailed  in \cite{PathIntegral}.  The same equations are found in \cite{ECFL}  working directly with the projected Fermi operators and the Heisenberg equations of motion.

By implementing a series expansion of terms in $\lambda$  we calculate equations to the required order, after which we set $\lambda=1$ before numerical evaluation. An interested reader can find the equations to second order in $\lambda$ for several interesting systems in \cite{ECFLall}.

It should be pointed out that the role  of the parameter $\lambda$ parallels the role of $\frac{1}{S}$ in a quantum spin system with interacting spin-S particles. This implies that  there are  formal similarities  between the ECFL theory and a semiclassical expansion for quantum spin systems. In the latter  problem, the spin-S objects are replaced by standard (canonical)  Bosonic variables within a systematic  Dyson-Maleev  expansion. An account of this connection can be found in \cite{PathIntegral}.

{\bf \S} For implementing Step (IV), it  is helpful to first  review the definition of a few objects that are needed for calculating the Greens function and related objects. In particular, the form of the expression \disp{formofG} can be understood from simple arguments that we present now.

The Greens function in \disp{G1} can be written in a minimal notation as
\beq
G(1,2)= - \lll C(1) C^\dagger(2)(1-n(\bar{2})) \rrr \label{G3}
\eeq
where $1$ and $2$ stand for position, (imaginary) time and spin labels, the bar over 2 is to remind us that the spin is to be reversed relative to the other operators, and the double brackets are shorthand for (imaginary)  time ordering and taking a thermal average with the weight factor $e^{-\beta H}/{\cal Z}$.   We write \disp{G3}  as a sum of two terms
\beq
G(1,2)=- \lll C(1) C^\dagger(2) \rrr \left\{ 1- \frac{\lll C(1) C^\dagger(2) n(\bar{2}) \rrr}{\lll C(1) C^\dagger(2) \rrr} \right\}
\eeq 
We {\bf define}  a  an auxiliary Greens function
\beq
g(1,2)= - \lll C(1) C^\dagger(2) \rrr   \label{g}
\eeq
which is a useful mathematical construct, in terms of which we decompose
\beq
\lll C(1) C^\dagger(2) n(\bar{2}) \rrr= - g(1,2) \frac{n}{2} + \lll C(1) C^\dagger(2)\widetilde{ n(\bar{2})} \rrr
\eeq
where $\frac{n}{2}$ is the average and  $\widetilde{ n(\bar{2})}= n(\bar{2}) -\langle \frac{n}{2}\rangle $ is the fluctuation of the operator $n(\bar{2})$. We further write
\beq
 \lll C(1) C^\dagger(2)\widetilde{ n(\bar{2})} \rrr= \lll C(1) C^\dagger(2)\rrr \times \Psi(1,2)
\eeq
where $\Psi$ is a dimensionless object that is reminiscent of a (dimensionless)  self-energy,  when defined through the Dyson  relation
 $i \omega_n G = \varepsilon G + \Psi G$,  arising for a Hubbard-type model with a dimensionless interaction term. For this reason we call $\Psi$ as the second self energy.  This is distinguished from the self energy $\Phi$ of the auxiliary $g$. Putting these together we get
\beq
G &=& g \times \{1-\frac{n}{2} + \Psi\} \label{Self-Psi} \\
 g&=& \frac{1}{G_0^{-1} - \Phi' } \label{Self-Phi}
\eeq  
so that combining these we obtain \disp{formofG}.

Returning to the equations of motion in \disp{Fermidentity}, we can express $\Psi$ and $\Phi$ as a series in powers of $\lambda$. The two self energy form of G \disp{formofG} is a very useful one for understanding the resulting spectral functions.
The Greens function can be cast in the form of Dyson's equation after rearrangement. The Dyson self energy is then a combination of these self energies in a somewhat complicated expression (see Eq.~(13) in \cite{ECFL-DMFT}).  The ``naturally'' occurring pair of self energies $\Psi$ and $\Phi'$  in this theory, with a relatively simple dependence on density and temperature, join together  to  produce a variety of low energy scales, such as the ones mentioned in the summary Section.~\ref{Summary}.


\subsection{{\bf  Summary of relevant    Equations for the \tJ model}}
For completeness, we provide   the equations satisfied by the Greens function for the \tJ model in 2-d \cite{2dResults1,2dResults2,ECFL-Edward-Shastry,Comments-Equations} to ${\cal O}(\lambda^2)$. These equations   yield the results quoted below in Section. The C-programs used\cite{Mike} and the results for the resistivity and spectral functions \cite{Sam} (in Python and {\em Mathematica}) are  publicly  available.  Readers wanting a broad overview  may  skip this technical section.

We use the abbreviation $k \equiv (\vec{k}, i\omega_n)$. We next collect the answers below in terms of the two self energies $\Phi',\Psi$. We  explicitly display the power counting parameter $\lambda$ which is set to 1 at the end,   and  the Lagrange multiplier parameter $u_0$.
We write
\beq
G(k)&=& g(k) \times \{1- \lambda \frac{n}{2} + \lambda  \Psi_\lambda(k)\} \label{eq27}
\eeq
with
\beq
 g^{-1}(k)= g_0^{-1}(k) - \lambda \Phi_\lambda'(k) \label{eq28}
 \eeq
 where
 \beq
 g_0^{-1}(k)= i\omega_n+{ \bmu} -\utoo -   {\varepsilon_k} \label{eq29}
 \eeq
  and
  \beq
  \Phi_\lambda'(k)&=& \lambda Y_1(k) + \lambda \Phi_\lambda(k) \label{eq30} \\
  Y_1(k)&=& - \frac{n}{2} \varepsilon_k  \nn \\
  \Phi_\lambda(k)&=& \bigchi_\lambda(k)+\varepsilon_k \Psi_\lambda(k) \label{eq31}
  \eeq
  In these expressions $\Phi_\lambda, \bigchi_\lambda$ and $\Psi_\lambda$ vanish at $\lambda=0$ and have a series expansion given by
 \beq
 \Psi_\lambda &=& \Psi_{[0]} +\lambda \Psi_{[1]} +O(\lambda^2)\nn \\
 \bigchi_\lambda&=& \bigchi_{[0]} + \lambda \bigchi_{[1]}+O(\lambda^2). \label{lambda-square-1}
 \eeq
Recalling the explicit factor of $\lambda$ multiplying $\Psi_\lambda$ in \disp{eq27} and $\Phi'$ in \disp{eq28},  this expression in \disp{lambda-square-1} gives corrections to $G$ upto ${\cal O}(\lambda^2)$.
We note expressions for the expansion coefficients 
\beq
\Psi_{[0]}(k)&=&0 \nn \\
 \Psi_{[1]}(k) & =& -   \sum_{pq}(\varepsilon_p+\varepsilon_q-u_0+J_{k-p})g(p)g(q)g(p+q-k), \label{Psi1}
\eeq
and
\beq
\bigchi_{[0]}&=& - \sum_p g_p (\varepsilon_p -\utoo+ \frac{1}{2} J_{k-p})  \label{chi0} \\
\bigchi_{[1]}(k)&=&
 -  \sum_{pq} \left(\varepsilon_p+\varepsilon_q-u_0 +\frac{1}{2}(J_{k-p}+J_{k-q})\right)\times(\varepsilon_{p+q-k}-\utoo+  J_{q-k}) \nn \\
 && \times g(p)g(q) g(p+q-k) \label{chi1}
\eeq

The second order i.e.  ${\cal O}( \lambda^2)$ calculation can be done by putting $\lambda=1$, substituting the energy dispersion $\varepsilon_k$ and exchange energy $J_k$ (the fourier transform of $J_{ij}$ for any lattice. Assuming a non-magnetic state the two parameters $\bmu$ and $u_0$ are determined from the two sum-rules
\beq
\frac{n}{2}&= \sum_k G(k) e^{i \omega_n \eta} \label{SR-1}  \\
\frac{n}{2}&= \sum_k g(k) e^{i \omega_n \eta}  \label{SR-2} \\
\eeq
where $n$ is the number of electrons per site,   $\omega_n= \frac{\pi}{\beta}(2 n+1)$,   $k=(\vec{k}, i \omega_n)$ and $\eta=0^+$.

We note that the single-particle spectral function $A(\vec{k},\omega)$, which is of central importance in understanding much of the physics of the \tJ model, can be inferred from the Greens function $G(k)$  \disp{eq27}. We first express $G$  in real continuous frequencies $\omega$ using analytic continuation from Matsubara frequencies $i \omega_n$. This is followed by using 
\beq
A(\vec{k},\omega)= - \frac{1}{\pi} \Im G(\vec{k}, \omega+ i 0^+) \label{spectral-function}
\eeq



\section{Results from the ECFL equations in d=0,1 and $\infty$ \label{others}}

Here  I list some applications of the ECFL method to  well-known strong correlations problems where exact solutions are known from other methods.

\begin{itemize}
\item[\S] {\bf   d=0 Anderson Impurity model (AIM)}
 Early formulations of the  ECFL equation were used to analyze the AIM in \cite{ECFL-AIM}.  In this problem  the numerical renormalization group  method  solution  originally formulated  by K. G. Wilson  and coworkers \cite{NRG,NRG2} gives exact results that can be compared against . The spectral function of the impurity electron calculated with ECFL agrees fairly well. One of the unexpected features of the ECFL theory is that  $\rho_\Sigma= - \frac{1}{\pi} \Im \Sigma(\omega)$, ( i.e negative  imaginary part of self energy) has a strong asymmetry term adding to the Fermi liquid quadratic behaviour, i.e. $\rho_\Sigma(\omega)= +|c|\omega^2 - |d| \omega^3 $, with a sign giving longer lived particles $\omega>0$, as opposed to holes $\omega<0$.  On the other hand,  an analogous asymmetry calculated in weak coupling perturbation theory \cite{Zlatic} has the opposite sign of the cubic term. This sign has significance in the theory of thermopower in correlated materials, where the sign of the   Seebeck coefficient has a contribution from the sign of this coefficient \cite{Zitko,Sign-Thermopower}. Our subsequent  work in \cite{Zitko} using Wilson's NRG method throws light on the sign of the above  asymmetric term. The $U$-density plane turns out to have a curve demarcating the crossover between the two signs of the asymmetric term.

 \item[\S] {\bf   d=1 t-t'-J model non-Fermi liquid spectral function}
 ECFL theory was applied to
 the t-t’-J model in one dimension  and the results compared with the time  dependent density matrix renormalization group of S. White \cite{DMRG,ECFL-1d}.
 In this situation the ground state is not a Fermi liquid, due to  low energy singularities, which also applies to the ECFL theory. It is interesting to see that the comparison  is quite good. In particular away from the region of low $\omega,T$ the  renormalization group  yields considerable structure in the spectral function when we vary $t'/t$, which is well captured by the ECFL solution.

\item[\S] {\bf   d=$\infty$ U=$\infty$ Hubbard model}

The ECFL equations can be analysed in the large $d$ limit and yields a Greens function with  $\vec{k}$ independent self energy, which has a $\lambda$ expansion, with terms that are  essentially the $d$=$\infty$ limiting cases  of  equations \disp{eq27}-\disp{chi1} \cite{ECFL-InfiniteD}. In
\cite{ECFL-DMFT} we compared  the results of ECFL and those of the dynamical mean field theory \cite{DMFT}.  The $U$=$\infty$ limit of DMFT is in fact the \tJ model with $J$=0, so the comparison is meaningful.  For a hypercube in the limit of 
 $d$=$\infty$, with various values of density and T, the results of the two theories  turn out to be reasonably close for most parameters.

 \end{itemize}

\section{Results  in d=2  }

In Refs.\cite{2dResultsFirst,Mai-Shastry,2dResults1,2dResults2} we implemented ECFL theory equations outlined above (\disp{eq27}-\disp{SR-2}) to the  2-d t-t'-J model. 
The objective was to  explore the dependence of  the quasiparticle weight, the spectral functions, the Hall constant and the resistivity for different values of the density, temperature and  the  parameters of the Hamiltonian \disp{HtJ}.
Here I present a selection of our results which should give the reader an idea about the scope of the theory and nature of the solutions, and provide some context to  the many results published in  detailed work.
The results from ECFL have two general  characteristics which are  evident below: a reduced quasiparticle weight $Z\ll1$, and the emergence of (several)  low temperature scales $T_*\ll t/k_B$ in the thermal variation of different physically  measurable quantities.

\subsection{ \label{Z}  Reduced Quasiparticle weight  Z}  One of the characteristic results that follow from the ECFL equations for the \tJ model is a small quasiparticle weight $Z=\{1- \frac{\partial}{\partial \omega} \Sigma(\vec{k}_F, \omega)\vert_{\omega\to 0}\}^{-1}$.   At half filling   $\delta$=0,  a Mott insulator results and  the  quasiparticle weight  vanishes.
For hole densities $\delta\gssim 0$ (i.e. particle densities $n\lessim 1$)  we are close to half filling.  In most cases that belong to this regime  we find $Z\ll 1$.
Such a  low value of  $Z$  is in striking contrast to results from the  weak coupling Fermi liquid theory. In that description  $Z$ does not vanish at half filling, and a much larger   $Z\lessim1$ is found  at all densities. In ECFL,    the small quasiparticle weight  close to half filling implies that the underlying Fermi liquid  has a small weight, and the incoherent part of the spectrum carries most of the weight. This fact  significantly influences   several  results of the ECFL theory. For example  we find broad backgrounds to spectral functions and with peaks that are thermally sensitive- interpreted as the effect  of a  low emergent   temperature scale.
 
\begin{figure}[H]
\centering
{\includegraphics[width=.55\columnwidth]{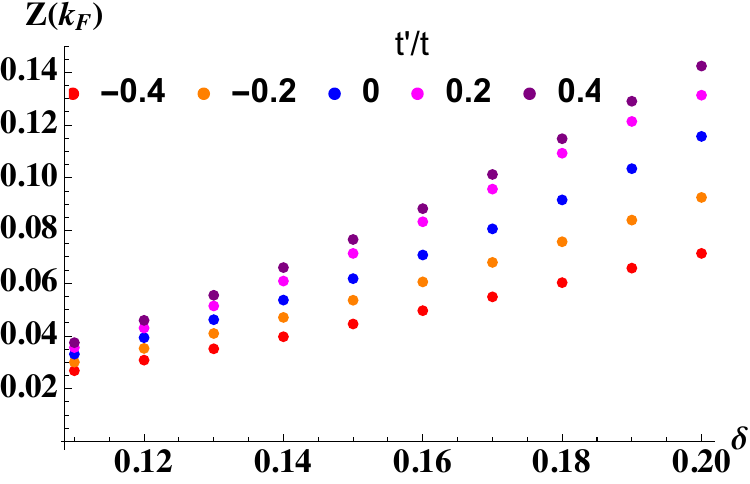}}
 \caption{\footnotesize The quasiparticle weight Z versus the hole density $\delta$ for different values of $t'/t$ within ECFL \cite{2dResultsFirst} showing that Z$\to$0 as $\delta$$\to$0. As a function of $t'/t$ it  is smallest at $t'/t=-0.4$ and increases towards $t'/t=0.4$.  { \it Adapted from \cite{2dResultsFirst} (CC BY 3.0)}
 \label{Zdelta} }
 \end{figure}
The calculated $Z$  vanishes at half filling, as seen in \figdisp{Zdelta} due to the Gutzwiller projection implicit in the theory. Its magnitude sensitively   depends  on the signs of the hopping parameters. The sign of the nearest neighbor $t$  always can be taken to be positive by a gauge transformation, but the sign of $t'/t$ cannot be freely chosen. For one thing, the shape of the Fermi surface depends crucially on this sign as well as its magnitude, and so does the curvature of the Fermi surface.  By convention, the electron doped class of \htsc systems are assigned $t'/t>0$ while the hole doped class is assigned $t'/t\leq0$. Our calculations for both signs of $t'/t$ are reported in \figdisp{Zdelta}. For hole doping we see that $Z$ drops to values of the ${\cal O}(\lessim  0.1)$ at densities $\delta \sim 0.15$, which correspond to the highest $T_c$ in \htsc  systems.  The electron doped cases $t'/t>0$ have a slightly larger value of $Z$ compared to the hole doped case, but also fall short of conventional Fermi liquid type Z values.

The dependence of $Z$  on the sign and magnitude of $t'/t$ in the theory is remarkable. Since different materials are described using different magnitudes and signs of these hopping, the application of the ECFL theory  leads to strongly material specific effects of correlations.
Its occurrence in theory may     {\em ex post facto} be  rationalized, as arising from the dual role of hopping  parameters in the\tJ model. The hopping parameters  not only   determine  kinetic energy  of Fermions, they   also enter the Hamiltonian  in a fashion that is analogous to interaction terms. This is  seen in the appearance of  $t_{lm}$'s as the coefficient of  a 4-Fermion term  in the Hamiltonian \disp{Hkin,HtJ2}.  Flipping the sign of $t'/t$  therefore changes several physical quantities in the ECFL theory in a way that is hard to predict {\em a priori}. With regard to the role of hopping parameters,  the \tJ model is structurally  different from Hubbard type models. In the latter, one can explicitly display  a clean separation between the role of the hopping parameters  and interactions \cite{Potthoff,deDominicis} within  a variational formulation of many-body theory\cite{deDominicis}. Here the hopping parameters $t_{lm}$
determine  $G_0$ the non-interacting Greens functions as expected, while only  interactions determine the  Luttinger-Ward functional $\hat{\Phi}[G]$,  which is  {\em independent}  of the hopping parameters. This object  is designed to  yield $\Sigma[G]$  from  $\frac{\delta \hat{\Phi}[G]}{\delta G}=\Sigma[G]$. The free energy  $\Omega$ is constructed as a functional of $G$:   $\Omega[G] =\hat{\Phi}[G]+ \Tr \ln G - \Tr G_0^{-1} G$.
Requiring stationarity,  i.e.  $\frac{\delta \Omega[G]}{\delta G}=0$ gives   Dyson's equation  rewritten as $G^{-1}+\Sigma=G_0^{-1}$, making it explicit that the hopping parameters entering through the  $G_0^{-1}$ term  serve as an initial condition for the renormalization of $G$.

\subsection{ \label{spectral-T}    Thermally sensitive    Spectral  Peaks,  Spectral functions and $\Im \Sigma(\vec{k},\omega)$}
\begin{figure}[thbp]
\begin{adjustwidth}{-1.5cm}{-1.5cm} 
\centering
{\includegraphics[width=.325\columnwidth]{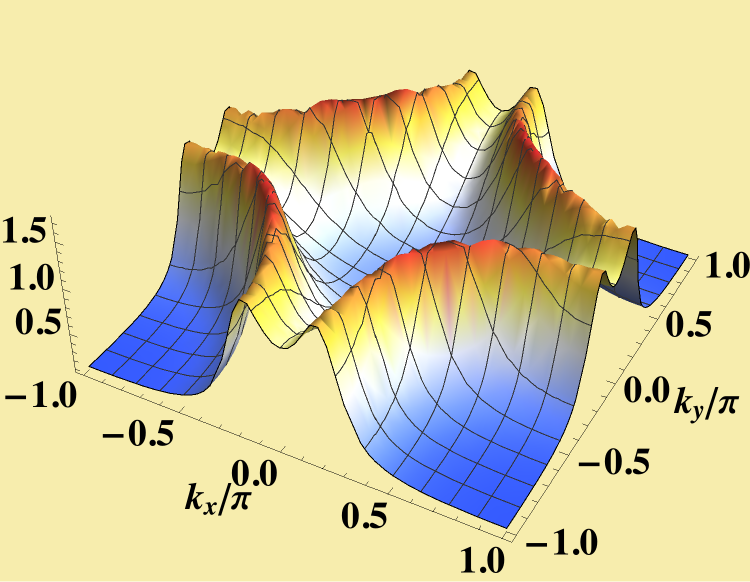}}
{\includegraphics[width=.325\columnwidth]{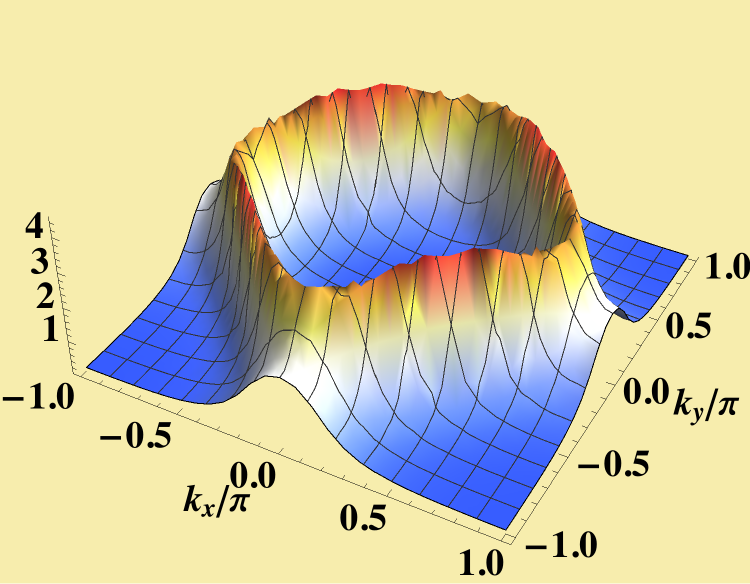}}
{\includegraphics[width=.325\columnwidth]{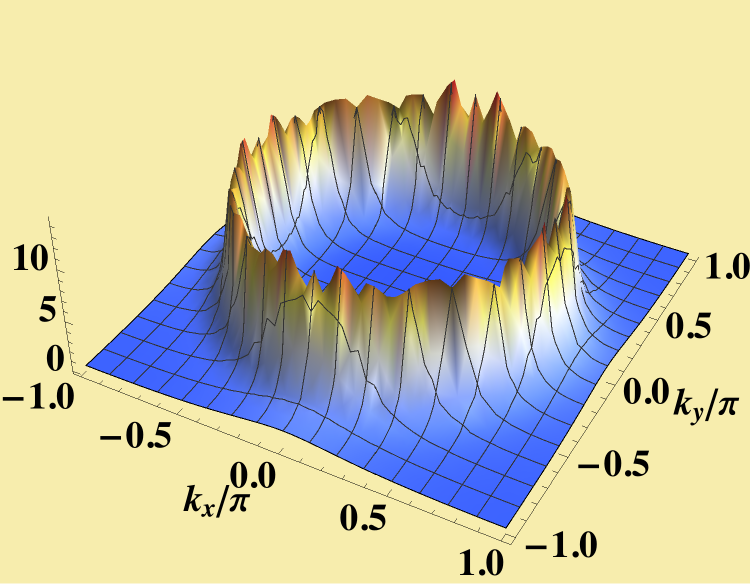}}
\caption{\footnotesize {\bf (L-R)} With  T=63 K and  hole concentration $\delta=0.15$, the $\vec{k}$ variation over the Brillouin zone of the maximal spectral peaks  $A_{max}(\vec{k})$ (i.e. $A(\vec{k},\omega_{peak})$) at three values of $t'/t$   from -0.4,0,0.4 . Observe that the relatively flat  peaks at -0.4 gives away to  steeper peaks at +0.4, accompanied by the change in curvature. The sign of the Hall constant $R_H$ has a contribution from this curvature, and within ECFL we find an electron  like $R_H$ for t'/t=0.4 and a hole like  $R_H$ for the other values, as seen in \figdisp{RHall}. {\it Adapted from \cite{{2dResultsFirst}}(CC BY 3.0)}
 \label{Peaks-1} }
 \end{adjustwidth}
 \end{figure}

\begin{figure}[htbp]
\begin{adjustwidth}{-1.5cm}{-1.5cm} 
\centering
{\includegraphics[width=.325\columnwidth]{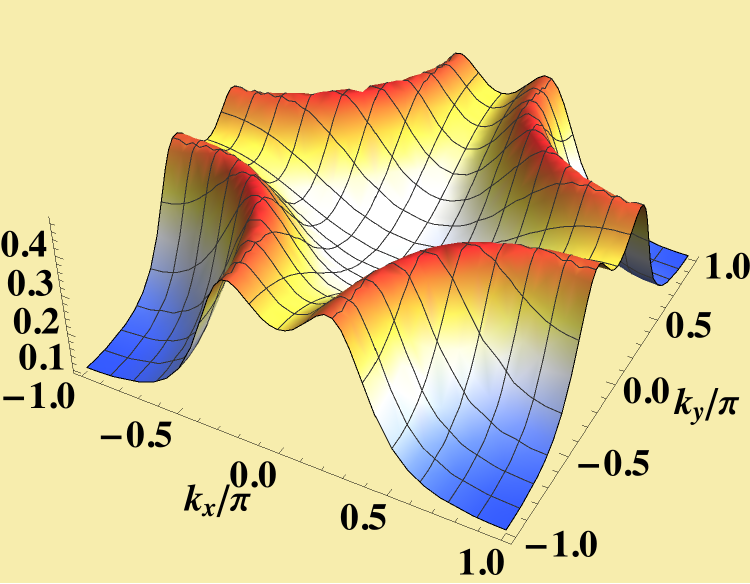}}
{\includegraphics[width=.325\columnwidth]{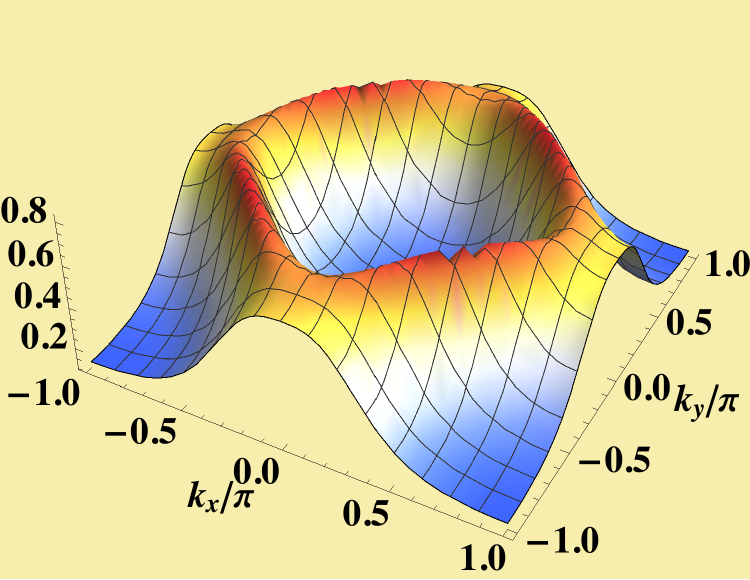}}
{\includegraphics[width=.325\columnwidth]{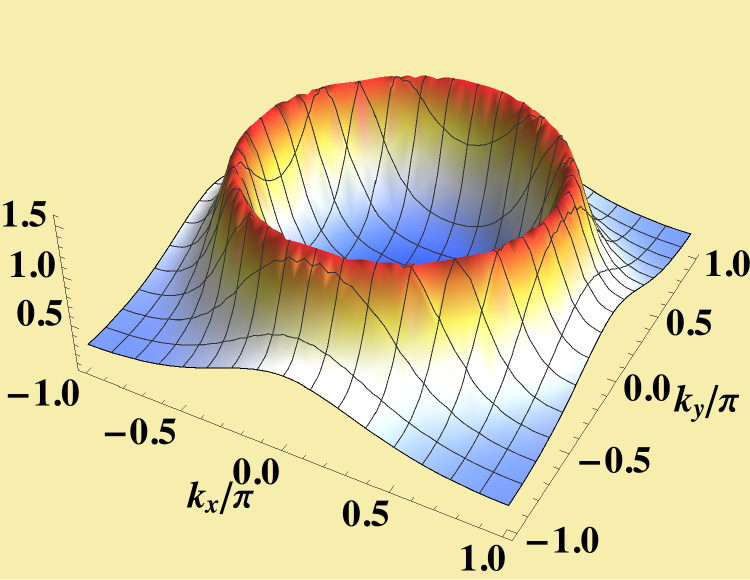}}
\caption{\footnotesize {\bf (L-R)} With a slightly higher  T=210 K and  $\delta=0.15$, the $\vec{k}$ variation over the Brillouin zone of the maximal spectral peak height (i.e. $A(\vec{k},\omega_{peak})$) at three values of $t'/t$   from -0.4,0,0.4. For the magnitude of the  hopping parameter t=5220K (i.e. bandwidth $\sim$ 3.45eV), the observed significant drop in magnitude of the peaks relative to those in \figdisp{Peaks-1} is remarkable. In a standard Fermi liquid  the corrections are ${\cal O}(T/T_F)^2$, and are expected to cause negligible drops for this (small) change of T. {\it Adapted from \cite{{2dResultsFirst}}(CC BY 3.0)}
 \label{Peaks-2} 
 \label{} }
  \end{adjustwidth}
\end{figure}

 We first discuss 
the spectral function $A(\vec{k},\omega)$, which  can be calculated from the electron Greens function using $A(\vec{k},\omega)= - \frac{1}{\pi} \Im G(\vec{k},\omega+ i 0^+)$.  As seen in \figdisp{EDC-1,EDC-2} ({\bf Left}), $A(\vec{k},\omega)$ displays peaks as a function of $\omega$ for a given  $k$$\sim$$k_F$ at  a characteristic energy $\hbar \omega_{peak}(k)$. The curve $\hbar\omega_{peak}(k)$ vs $k$ defines the dispersion relation discussed further below.  We first discuss the  parameter dependence of spectral peaks within the theory. By this we mean  the  maximal spectral functions $A_{max}(\vec{k})=A(\vec{k},\omega_{peak}(k))$.
We calculate the spectral peaks at a fixed density over the entire zone, using  $t'/t$=-0.4,0.,0.4 in \figdisp{Peaks-1,Peaks-2}. These values of $t'/t$    span the range from electron doped to hole doped systems. These calculations were performed with the hole density $\delta$=0.15 and the bare electronic band-width $\sim 3.5$ eV.  The    results  at T=63K in \figdisp{Peaks-1} display the change in topology of the Fermi surface from hole-like (i.e. open) to electron-like (closed). We   also see  a sharpening of the peaks, i.e. the magnitude of the peak increases by almost an order of magnitude as $t'/t$ varies from -0.4 to 0.4. The inelastic  background  also decreases correspondingly, since the total weight of $A(\vec{k},\omega)$  (in the lower Hubbard band) is conserved thanks to  a parameter independent sum-rule $1-\frac{n}{2}$. 

In \figdisp{Peaks-2} calculated at T=210K, a slightly higher temperature given  the scale of the band-width, we observe a substantial  flattening of the peaks by  a factor varying between   $\frac{1}{5}$ and $\frac{1}{10}$.  This is an indicator of a low characteristic  temperature scale at play. As remarked above, this thermal sensitivity  is a hallmark of the ECFL theory. This type of sensitive thermal variation is also seen in the resistivity and Hall constant discussed below.

\begin{figure}[thbp]
\begin{center}
\includegraphics[width=0.55\columnwidth]{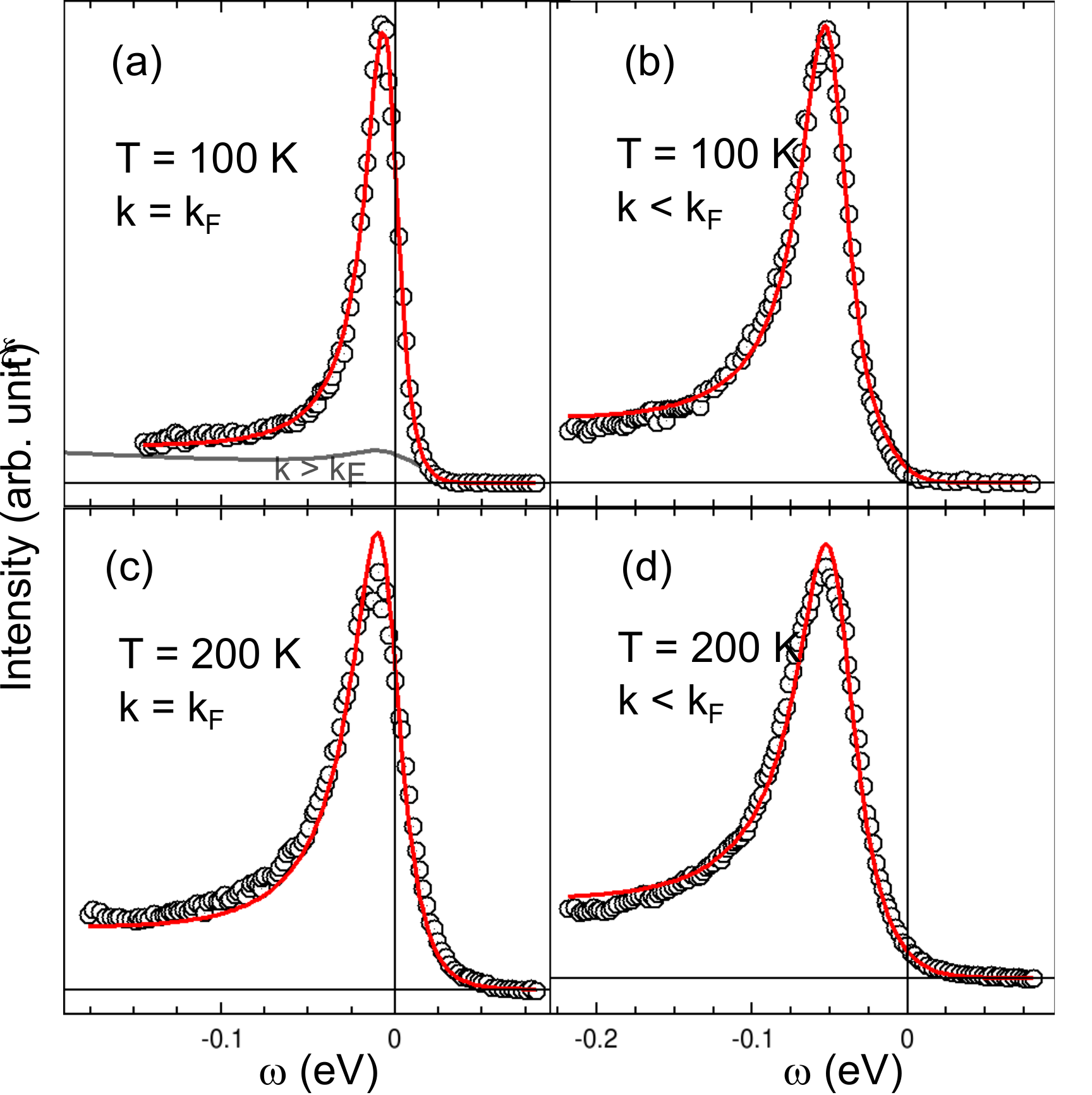}
\caption{ \footnotesize The spectral intensity  $I\propto A(k,\omega)$ multiplied by the Fermi function $f(\omega)$, from experiments \cite{casey}  at $\delta$=0.15 (symbols), is fit with a phenomenological line shape (red lines) obtained from \disp{eq27} \cite{Gweon}. Microscopic calculations of the line shapes   for several sets of parameters can be found in \figdisp{EDC-1,EDC-2}. {\it Adapted from \cite{Gweon},\copyright(2011),American Physical Society}.
	\label{Spectral-Functions}}
\end{center}
\end{figure}

\begin{figure}[bthp]
\begin{adjustwidth}{-1.5cm}{-1.5cm} 
\centering
{\includegraphics[width=.55\columnwidth]{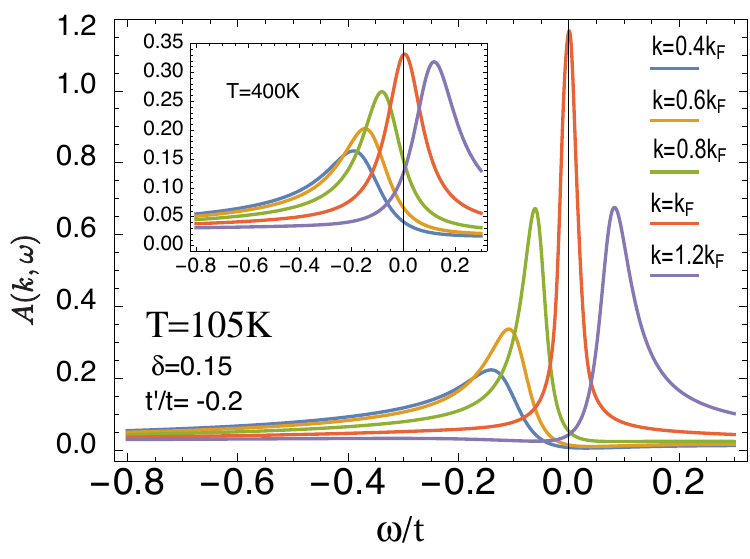}}
{\includegraphics[width=.55\columnwidth]{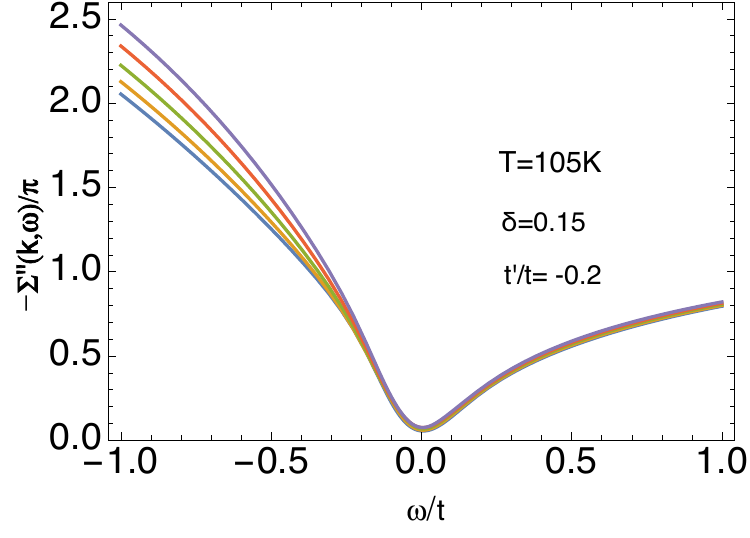}}
 \caption{\footnotesize Hole doping with  J/t=0.17and  t= 0.45eV  from \cite{Mai-Shastry} {\bf Left:} The $\vec{k}$ and T dependence of the  spectral function $A(\vec{k},\omega)$   along the $\{0,0\} \to\{\pi,\pi\}$ direction.  \label{EDC-1}  The temperature difference between the main figure and inset  here and in \figdisp{EDC-2}, is  very small on the scale of the bandwidth $\sim$3.6$\times10^4 $K.
  The substantial reduction of the magnitude    indicates  the high  thermal sensitivity  of the ECFL spectra.  {\bf Right:}   The main figure and the inset display the imaginary part of self energy using the same parameters. {\it Adapted from \cite{Mai-Shastry},\copyright(2018),American Physical Society}.}
\end{adjustwidth}
 \end{figure}

\begin{figure}[bthp]
\begin{adjustwidth}{-1.5cm}{-1.5cm} 
\centering
{\includegraphics[width=.55\columnwidth]{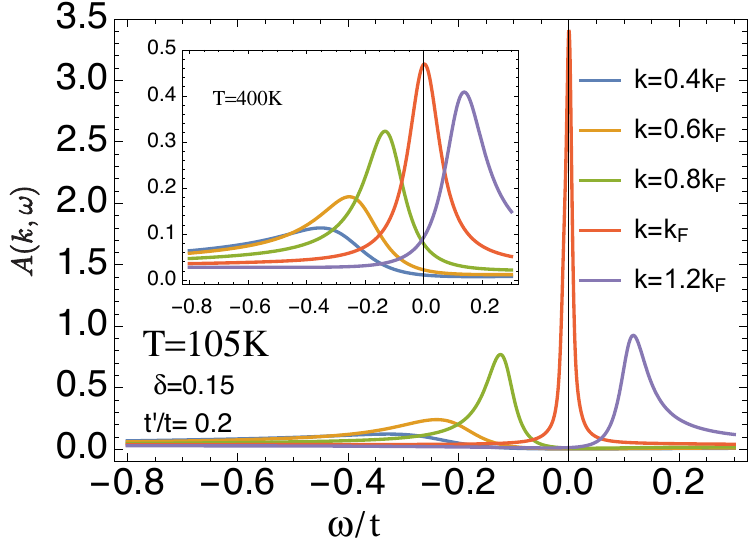}}
{\includegraphics[width=.55\columnwidth]{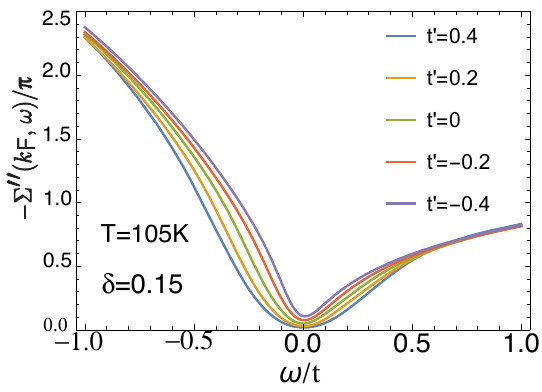}}
 \caption{\footnotesize With  t'/t=0.2 (i.e. electron doping)  and with  otherwise identical   parameters as  in the hole doped case  \figdisp{EDC-1}. {\bf Left:} The $\vec{k}$ and T dependence of the  spectral function $A(\vec{k},\omega)$    along the  $\{0,0\} \to \{\pi,\pi\}$ direction. Notice the greater sharpness and  peak heights relative to hole doping.  {\bf Right:}  The imaginary part of self energy at different values of t'/t.  As t'/t reduces from 0.4 to -0.4,  the  $\omega^2$ behavious  near $\omega$$\sim0$ is more pronounced. \label{EDC-2}. {\it Adapted from \cite{Mai-Shastry},\copyright(2018), American Physical Society}. }
\end{adjustwidth}
 \end{figure}

The spectral functions calculated in ECFL  has a strongly asymmetric  shape that differs significantly from that of Fermi liquids. In \figdisp{Spectral-Functions} we display very early results from the ECFL theory in 2011 \cite{Gweon}, found from the expressions in \disp{eq27}, using a phenomenological parametrization of  the two self energies $\Psi,\Phi$. These match  well the unusually shaped Laser ARPES spectra that were  becoming available at that time \cite{casey}. Later microscopic calculations in ECFL \cite{2dResultsFirst} (Fig.1,Fig.2), and in \figdisp{EDC-1,EDC-2} below,  provide  an elaboration  of these line shapes. These illustrate the line shapes in different directions in the Brillouin zone, and their  complex evolution with the band parameters,  T and density. In \figdisp{EDC-1,EDC-2} we also show the calculated frequency dependence of  $\Im \Sigma$, and its dependence on T as well as the band parameters.

\subsection{\label{Kinks} Dispersion relation kinks}
The interesting feature of ``kinks'' is seen in the spectral function found in ECFL \cite{Gweon}, and also in several experiments. The kink   locates an abrupt  change in slope in the dispersion curve, i.e. the $\omega_{peak}(\vec{k})$ versus $\vec{k}$ curve. In brief the kink originates in the non-Lorentzian  line shape obtained from \disp{eq27}, and does not exist in a purely Lorentzian spectrum.  It is known that electron-phonon interactions  also lead to kink that are superficially similar to the ones found from correlations in ECFL. We discuss below a possible method to distinguish between the two mechanisms.

\begin{figure}[thbp]
\begin{center}
\includegraphics[width=0.45\columnwidth]{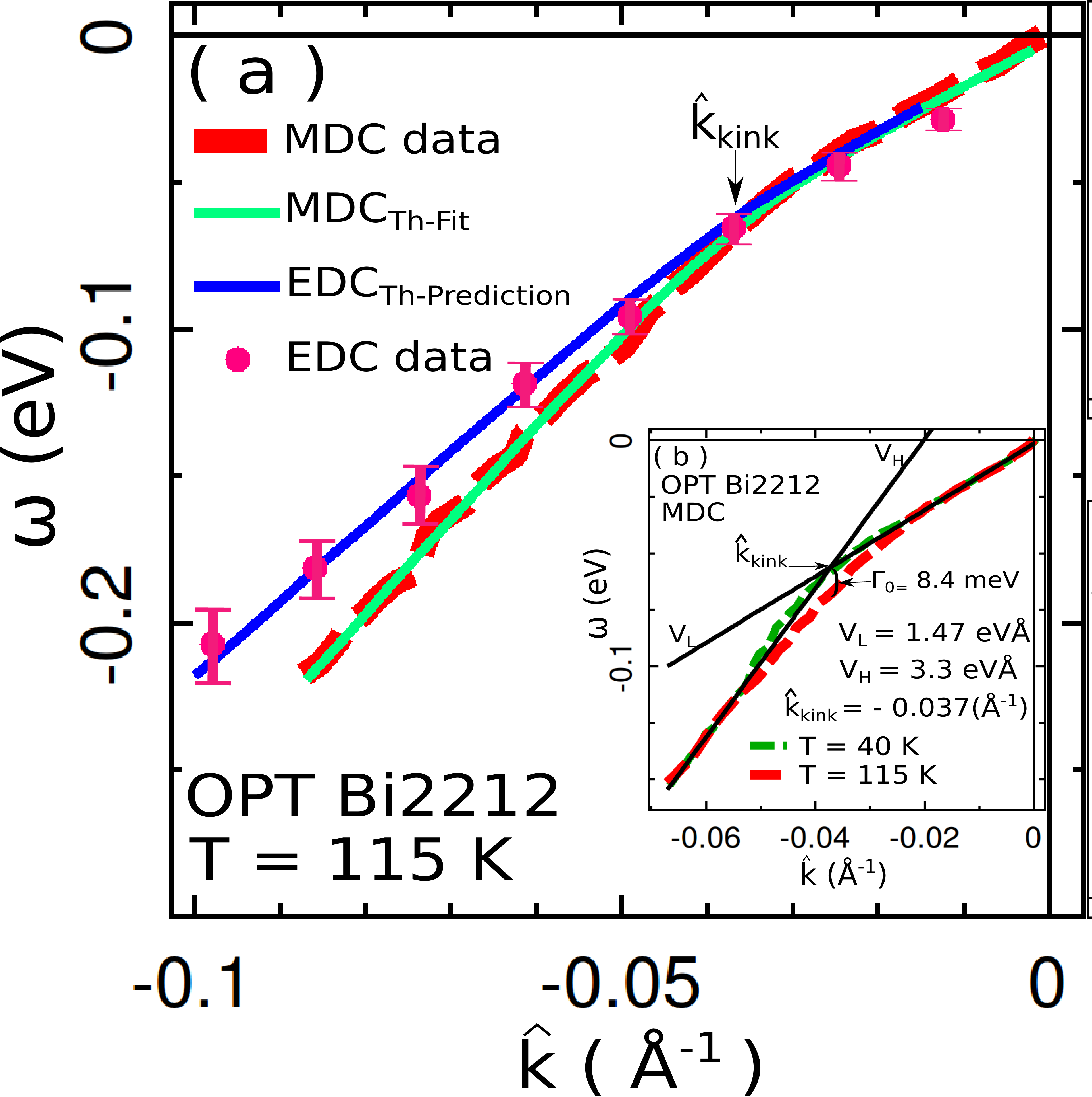}
\caption{\footnotesize  Fig. {\bf (a)} shows the dispersion for OPT Bi2212 at T=115 K \cite{Kaminsky}
from both EDC and MDC methods, and locates a kink at the indicated location with $E_{kink}$$\sim$65 meV and $\hat{k}$$\sim$-0.037$\ang^{-1}$. The blue line is the prediction for the $V_H^*$ obtained from \disp{kinks}, overlaid by the EDC data. The inset {\bf(b) } shows the variation of the kink with temperature, which is discussed further in \cite{Kazue}. {\it Adapted from \cite{Kazue},\copyright(2017),American Physical Society}.
	\label{Fig-Kinks}}
\end{center}
\end{figure}

 Denoting as $\hat{k}$ the deviation of $\vec{k}$ from $\vec{k}_F$, i.e. $\vec{k}$-$\vec{k}_F$ along some direction in the Brillouin zone, and $\omega$ as the binding energy, the dispersion curves are typically linear in  the momentum  as seen in \figdisp{Fig-Kinks}, and therefore may be represented by a velocity $V$. The energy dispersion can be inferred either  from taking  fixed $\omega$ sections of $A(\vec{k},\omega)$, called  the momentum distribution curves (MDC's), or from taking fixed $\vec{k}$ sections,  termed as the energy distribution curves (EDC's).  These  dispersions are  different in general, due to the non-Lorentzian line shapes \cite{Kazue}.  We may define two distinct  velocities $V$ from the MDC's and $V^*$ from the EDC's.
  
  Kinks are seen in both EDC and MDC dispersions theoretically and also experimentally. 
 Close to $\vec{k}_F$ the velocities $V^*$ from  EDC dispersions and $V$ from MDC  are identical- call them $V_L$. For $\hat{k}$ beyond a certain kink wave vector  $\hat{k}_{kink}$, the EDC velocity $V_H^*$ and MDC velocity $V_H$ differ in magnitude. These   three distinct velocities can be seen in \figdisp{Fig-Kinks}, where the data is taken from  optimally doped Bi2212 system \cite{Kaminsky}.


  The ECFL theory line shape  noted  in \cite{Kazue} predicts that of the three measurable velocities $V_L$, $V_H$ and $V_H^*$, only two are independent. They are related through the formula 
\beq
V_H^*= \frac{3 V_H-V_L}{V_H+V_L} \times V_L ,\label{kinks}
\eeq
which gives $V_H^*$ in terms of the other two.  Within  the electron-phonon mechanism there  is no  reason for   a relation like \disp{kinks}  to exist.

This remarkably simple relation seems to be well satisfied in the Bi2212 data as seen in \figdisp{Fig-Kinks}. It also seems to hold in some further data sets examined in \cite{Kazue}.  A more thorough re-investigation of  data on  kinks in correlated systems, informed by these insights, seems worthwhile.

\section{Transport results from ECFL \label{Transport}}

We first summarize the relevant formulas used to calculate the resistivity and the  Hall response from the \tJ model using the ECFL formalism.
 Within the ECFL theory, the longitudinal resistivity $\rho_{xx}$ can be obtained by  taking  2-d layers of \tJ model in the x-y plane with  lattice constant $a_0$, stacked on top of each other with a separation $c_0\gg a_0$. The constant $c_0$ is related in a simple way to the c-axis lattice constant. In this picture the electrical  conduction is assumed to be in the plane of the 2-d layers acting in parallel, and yields the component $\sigma_{xx}$  of the conductivity. The resistivity   is calculated from   the computed spectral weight $A(\vec{k},\omega)$ \disp{spectral-function} using  the formula \cite{2dResultsFirst,2dResults1,2dResults2} for the scaled (dimensionless) resistivity $\bar{\rho}$
\beq
\frac{1}{\bar{\rho}}= \frac{(2 \pi)^2}{a_0^2} \int_{-\infty}^\infty d\omega \, (-\frac{\partial f(\omega)}{\partial \omega}) \langle A^2(\vec{k},\omega) (\hbar v_{k}^x)^2\rangle_{\vec{k}} \label{rho-formula}
\eeq
where $\vec{v}_k=\frac{1}{\hbar} \nabla_k \varepsilon_{\vec{k}}$ is the group velocity of electrons in the non-interacting  band, $f(\omega)=\{e^{\beta \omega}+1\}^{-1}$ is the Fermi function, and the angular bracket  indicates a normalized integral over all $\vec{k}$.  From the  scaled resistivity, we obtain the  longitudinal resistivity $\rho_{xx}$ using 
\beq
\rho_{xx}= \frac{h}{e^2}\times c_0\times \bar{\rho}\left(\frac{t'}{t},\frac{t''}{t},\frac{k_B T}{t}, \frac{J}{t},n \right) \label{scaled-resistivity}.
\eeq
The argument of $\bar{\rho}$ lists all the  parameters of the Hamiltonian \disp{HtJ}.  In addition to the density $n$ (often denoted below by the hole density $\delta$=1-$n$), the other variables are   the hopping parameters, the exchange J, and temperature T scaled by the nearest neighbour hopping t. This formula  \cite{2dResults1,2dResults2,2dResultsFirst} omits vertex corrections. From this formula we obtain the absolute value of the resistivity, which can be compared to data. 

 Within this  scheme,  we may also calculate the Hall conductivity\cite{voruganti,Tremblay,2dResultsFirst} as $ \sigma_{xy}= -  \frac{2 \pi^2 e^2}{h c_0} \times (\frac{\Phi}{\Phi_0}) \times \; \bar{\sigma}_{xy} $,   the scaled (dimensionless) conductivity:
  \beq 
 \bar{\sigma}_{xy}&=& \frac{4 \pi^2 }{3 }   \int_{-\infty}^\infty d\omega \, (- {\partial f}/{\partial \omega}) \langle A^3(\vec{k},\omega) \eta(k) \rangle_k, \;\;\;\; \label{hall1}
 \eeq
 and 
 \beq
 \eta(k)={ \frac{ \hbar^2}{  a_0^4}} \{ (v_k^x)^2  \frac{\partial^2 \varepsilon_k}{\partial k_y^2}- (v_k^x v_k^y) \frac{\partial^2 \varepsilon_k}{\partial k_x \partial k_y} \}. \label{Hall-eta}
 \eeq 
 The dependence of $\bar{\sigma}_{xy}$ on the band parameters is similar to that of $\bar{\rho}$ in \disp{scaled-resistivity}. 
 In terms of these we can compute the  Hall constant $R_H$ and   Hall angle $\Theta_H$   as
 \beq
 c \, R_H & =& - \frac{4\pi^2 c_0 a^2_0}{|e|}  \; \bar{\sigma}_{xy} \times \bar{\rho}_{xx}^2, \label{dimensionless-Hall} \\
  \cot(\Theta_H)&=& - \frac{1}{2 \pi^2}  \frac{\bar{\sigma}_{xx}}{\bar{\sigma}_{xy}} \times \frac{\Phi_0}{\Phi}. \label{cot-Hall}
 \eeq
  Here    $\Phi=B a_0^2$ is  the  flux   and $\Phi_0= hc/(2 |e|)$ is the flux  quantum.

\subsection{\label{Hall} The  Hall constant and Hall angle:}
\begin{figure}[htbp]
\begin{adjustwidth}{-1.5cm}{-1.5cm}
\centering
{\includegraphics[width=.55\columnwidth]{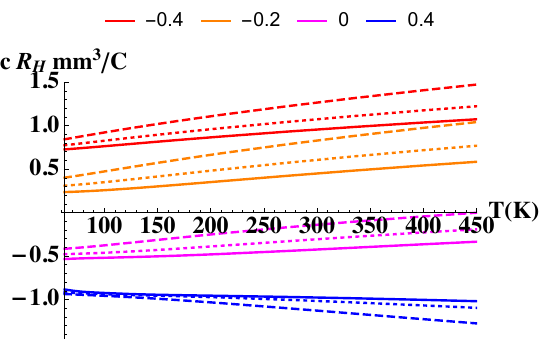}}
{\includegraphics[width=.55\columnwidth]{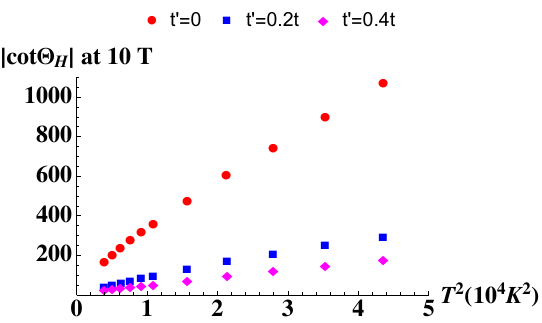}}
 \caption{\footnotesize {\bf Left} The calculated  Hall constant $R_H$ at $\delta$=0.15 for different values and signs  of the second neighbour hopping $t'/t$. The change in sign is consistent with the change in topology of the Fermi surface seen in \figdisp{Peaks-1,Peaks-2}.  {\bf Right} The calculated cotangent  Hall angle shows a linearity with $T^2$, of the type seen in  experiments \cite{Takagi,Ong}. We used $t$=0.45 eV, $a=3.79\, \AA^0$ and $c_0=6.65 \, \AA^0$, $v_0/|e|= .596\times10^{-3} \,cm^3/C$   and  $\Phi_0/\Phi= 1440$ with $B=10T$. {\it Adapted from \cite{2dResultsFirst},\copyright(2018), (CC BY 3.0)}.
 \label{RHall} }
 \end{adjustwidth}
 \end{figure}
The Hall constant $R_H$, the Hall angle  and other measures of magneto-transport can be calculated within ECFL \cite{2dResultsFirst} using \disp{dimensionless-Hall,cot-Hall}. In \figdisp{RHall}  ({\bf Left}) we show the T  dependent Hall constant at a density $\delta$=0.15 calculated  with various values of $t'/t$. It is noteworthy that the sign of $R_H$ is electron-like (i.e. negative) for $t'/t$=0, 0.4, while it is hole-like (i.e. positive) for $t'/t$=-0.2,-0.4. This change of sign is consistent with the observation that electron and hole doped \htsc materials show opposite signs of  $R_H$.  Near the Mott insulating limit at high T, interactions are expected to play a role in  determining the sign of the Hall constant \cite{Shastry-Hall}. However at low T the   band structure determined  curvature effects  come into play strongly.  These are visible in  photoemission studies which show opposite signs of the  Fermi surface curvature between the two classes of systems. We comment further on the change in Fermi surface curvature below in \figdisp{Peaks-1,Peaks-2}.
The cotangent of the Hall angle plotted versus $T^2$ is  displayed in \figdisp{RHall}, and shows a linearity that was noted in experiments \cite{Takagi,Ong}.   The change in slope in \figdisp{RHall}({\bf Right}) is prominent for $t'$=0 for $T^2 \gssim$ $10^4$K,  and is also 
seen in the other curves. A more detailed study of the characteristic temperature scale  defined through the Hall angle and its dependence on the parameters t, and  t'/t is currently in preparation \cite{Samantha-Hall}. It is interesting to note that this  change in slope was already seen in the data \cite{Takagi,Ong}, but not remarked upon. This could  also be a useful direction for further detailed  experimental study.

\subsection{The Resistivity: \label{Resistivity}}
Understanding the observed quasi-linear $\rho(T)$ behaviour is  widely recognized as one of the major problems in understanding the physics of \htsc materials.  In experiments
within a family of materials with similar composition \cite{Names}, there are two kinds of systematics that need to be simultaneously understood.
At a fixed density and over a wide range of temperatures, one observes different  regimes exhibiting distinct types of  $\rho(T)$ vs $T$  variations.
 As the  density is changed,   the magnitude of   resistivity and  its distinctive T-dependence  regimes   also change. The measurement  of $\rho(T)$  over a wide range of T and densities poses the greatest theoretical challenge, and therefore   is  of most interest. 
 
 In this   extended section I present the  resistivities from the ECFL theory, focussing on   their T and density   dependence  using different sets of theoretical parameters and compare with available data.   Starting  from a 2-d \tJ model within ECFL theory in \cite{2dResults1,2dResults2,2dResultsFirst},   we calculated  the resistivity of all available   single layer \htsc materials \cite{Names}(also see Table-1), using formulas given in \disp{scaled-resistivity}.   We focussed on the class of single layered systems, which require the least number of theoretical assumptions.   Two and higher  layer systems involve making extra  theoretical assumptions about the coupling between layers, and therefore are avoided here. We stress that the class of single layer materials have considerable variety in their band hopping parameters- determined most directly from photoemission studied (see Table-1).  Addressing the visibly different effects of correlations within this class of materials provides a non-trivial challenge to the theory.

The experimental data often requires to be adjusted for eliminating the effect of disorder. The disorder  is assumed  small enough for an additive correction to suffice. Some recent experiments quote their results after removing such an additive disorder contribution \cite{LCCO,Cooper,Hussey}, so that in these cases such an adjustment is not necessary.

In order to start from a fully determined model, we need to fix  the  values of the  
 hopping parameters $t,t'$ (sometimes $t''$ as well),   for each class of systems. 
   Of these parameters $t'/t$  can in all cases studied here,  be fixed  from the ARPES determined Fermi surfaces (FS) of these compounds quite accurately.
 From our calculations with varying values of $J$ we find our results are not sensitive to this parameter in the density range of interest, and we chose $J/t=0.17$ for all our studies.
 
  We  fix the only remaining parameter $t$ from experiments, at  {\em one density}, typically chosen to be hole  density $\sim$0.15, and a  temperature $T^{\Phi}$ that is medial to the experiment. At this density we take the experimentally measured slope $\Gamma(T^{\small \Phi})=\frac{d \rho}{d T}\vert_{T^{\Phi}}$, and match it with the theoretical results by varying the theoretical $t$. This set of parameters, $t,t',t'',..J$ are then used for the entire data sets at all available densities for that particular compound.

We  applied this protocol to  all known single layer \htsc systems   listed below in Table-I, where the number of distinct measured samples $N_{samp}$ is given in the second column.   In some cases the determination of t'/t from the Fermi surface is not unique, and more than one set of hopping parameters give equally good fits.  In   Table-I for the \htsc material Tl2201, we show {\em two} sets of such parameters, which are quite distinct. We calculated the resistivity with both sets of parameters, which give almost identical resistivities as discussed below. 
\begin{table}[tbhp]
\centering
\begin{tabular}{||p{3.7cm} |c|c|c|c|c||}
\multicolumn{6}{c}{{\bf  All Single Layer Cuprate High $T_c$ Materials}}  \\ \hline \hline
{\bf  Hole-doped } &x-range ($N_{samp}$)&$T_{max}$(K)&$t'/t$& $t''/t$  & $t$ (eV)   \\ \hline \hline
 $La_{2-x}Sr_xCuO_4$ (LSCO)\; \cite{Ando-1} &0.12-0.22 (11)&400& -0.2  &0& 0.9   \\ \hline
 $Bi_{2}Sr_{2-x}La_xCuO_6$  (BSLCO) \; \cite{Ando-1}&0.12-0.18 (7)&300& -0.25  & 0&1.35  \\ \hline 
 $Bi_2Sr_2CuO_{6+x}$ (Bi2201)  \cite{Hussey} &0.213-0.258 (4)&300&  -0.4   & 0 & 1.176 \\  
(Bi2201)  \cite{Fiory,Martin} &0.259\{0.32?\} (1)&800&-0.4&0&1.176    \\ \hline
 $Tl_2Sr_2CuO_{6+x}$ (Tl2201ModelA) \cite{Cooper2}&0.183-0.274 (4)&300& -0.430 & 0.005 &1.82  \\  
 (Tl2201ModelB) \cite{Cooper2}&&& -0.237 & 0.138 &1.053 \\ \hline 
 $HgBa_2CuO_{4+x}$ (Hg1201) \; \cite{Yamamoto}&0.127-0.208 (4)&300& -0.228 & 0.174 & 0.22  \\ \hline \hline
{  \bf Electron-doped } &&&&   &    \\ \hline \hline
 $Nd_{2-x}Ce_xCuO_4$ (NCCO) \; \cite{NCCO} &.125-.15 (2)&400& +0.2  &0& 0.9   \\ 
 $La_{2-x}Ce_xCuO_4$ (LCCO)\; \cite{LCCO}&.14-.17 (4)&300& +0.2 & 0&0.76   \\ \hline
 \hline 
\end{tabular}
 \caption{{\footnotesize The second column gives the range of hole density $x$ (i.e. $\delta$) and the number of samples available in that range. The first five rows consist of the known hole doped single layer materials and the last two are the electron doped single layer materials. 
 The highest  measured temperature   is  the  third column.  In the last three columns we list  the band parameters used in the theory.} } 
\label{All-Compounds}
\end{table}

\newpage

We next show a selection of  calculated resistivities overlaid with experimental data,  the full set of results for all single layer cuprate materials\cite{Names}  can be found in   \cite{2dResults1,2dResults2}.
 In our  discussion we noted  that  amongst the family of \htsc systems, those with  $\rho$-vs-$T$ data available  for a large number of  densities provide the most stringent tests.  Almost all the data sets have similar temperature ranges, while  LSCO and BSLCO have a large number of available samples (11 and 7) at well defined densities, and following the discussion above they are  particularly interesting.  For   LSCO and  BSLCO  presented below, we digitized the published data using Digizelt software.  This digitizing process turned out to have some limitations on acquiring the low-T data due to overlapping curves from nearby densities, leading to a curtailment of the lower end of the displayed T range in certain figures.


\begin{figure}[H]
\begin{adjustwidth}{-1.5cm}{-1.5cm} 
\centering
{\includegraphics[width=.45\columnwidth]{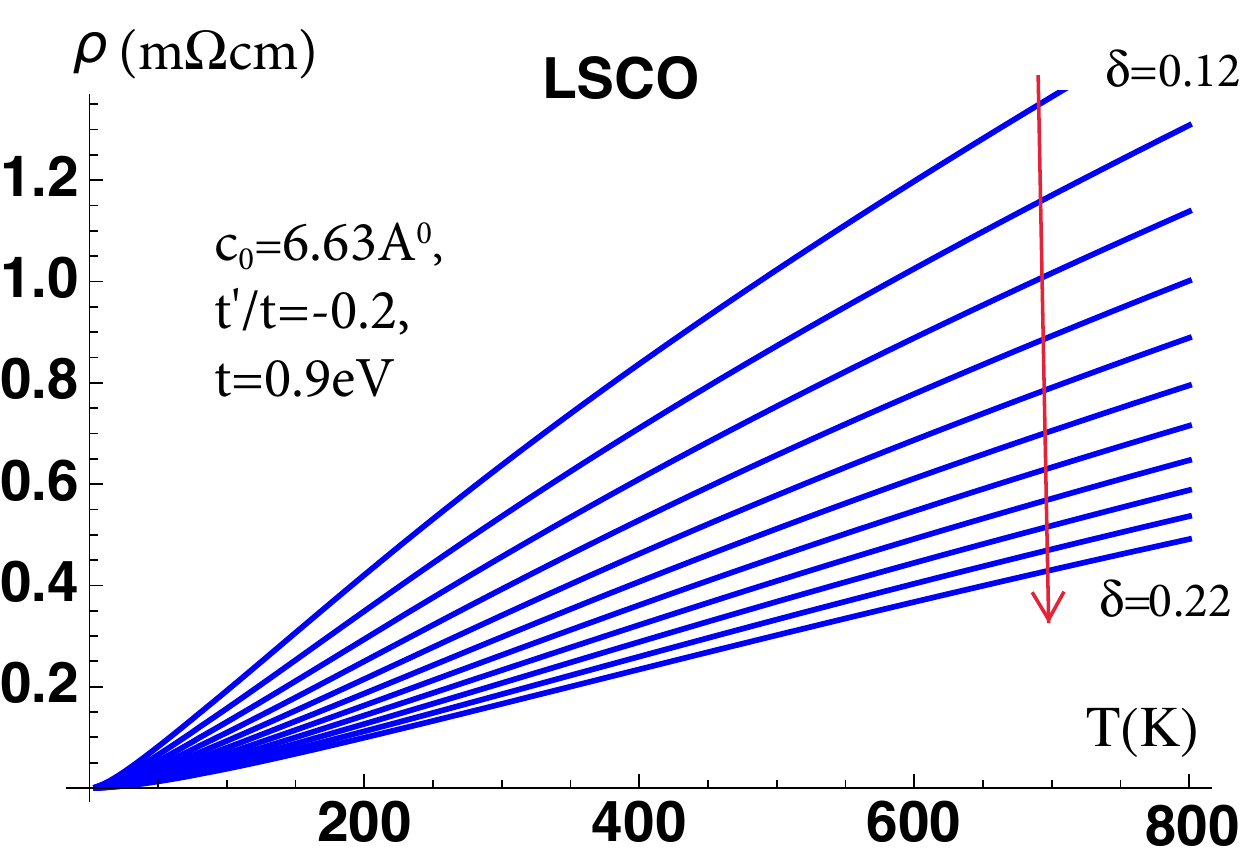}}
{\includegraphics[width=.45\columnwidth]{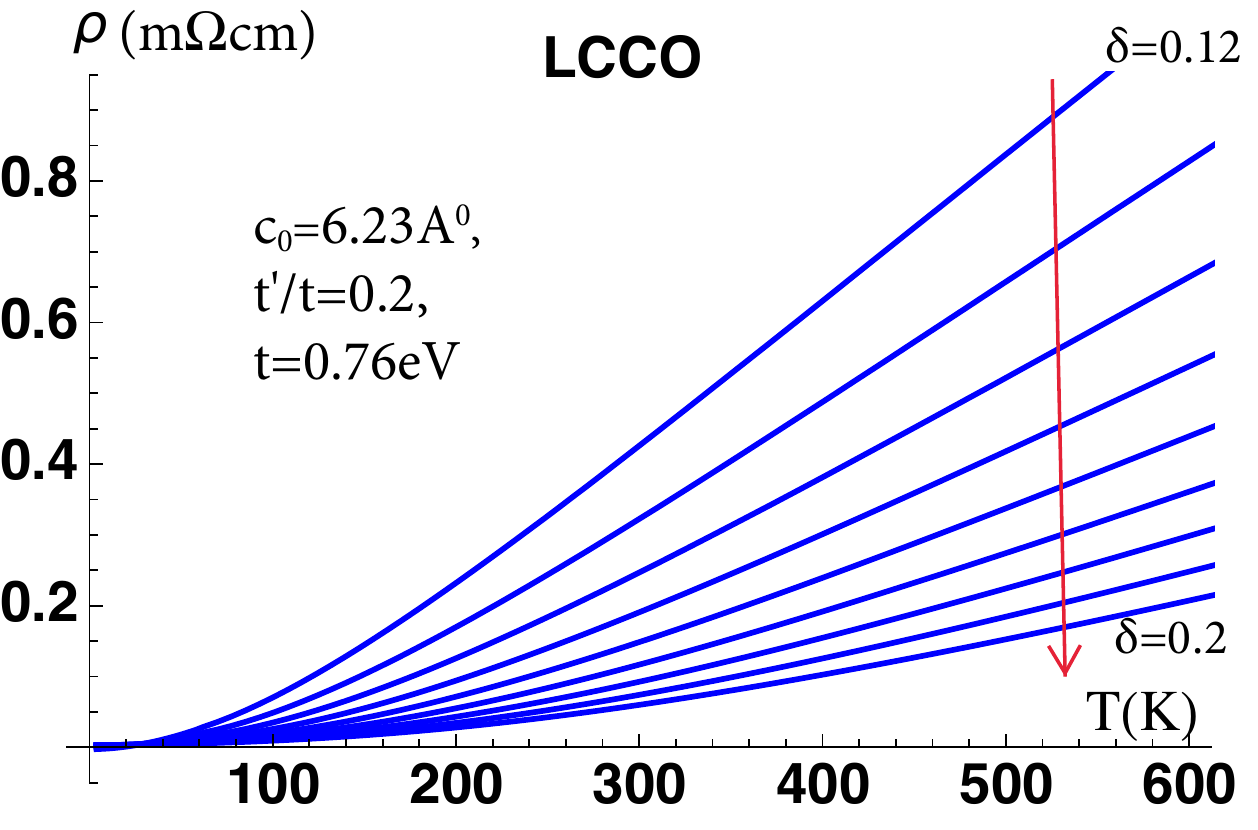}}
 \caption{\footnotesize {\bf  ECFL  resistivities over a wide T range:} Resistivity of LSCO, BSLCO and LCCO computed from ECFL \cite{2dResults1} over a wide range of densities and T.  
 At low T$\lessim$100K all curves exhibit $\rho(T)$$\sim$$T^2$ behaviour. Overall 
  LCCO with  $t'/t>0$ exhibits  a enhanced regions showing  $\rho(T)$$\sim$$T^2$  behaviour  compared to the hole doped  LSCO and where $t'/t<0$.
LSCO  exhibits  a low T crossover from $T^2$ to  a quasi-linear region with   $\rho(T)$$\sim$$T$. For low $\delta$   the resistivity    bends over like $\rho$$\sim$A-B $T^2$ ( with B$T^2\ll$A), while the large $\delta$$\gssim$.16 regime is  quasilinear over a wide temperature range. \label{All-T}. {\it Adapted from \cite{2dResults1},\copyright (2020) }}
\end{adjustwidth}
 \end{figure}
Let us first examine the theoretically computed resistivities  from  ECFL theory for LSCO and LCCO, presented  in \figdisp{All-T} over the T range  $0\leq T \leq 800$K at several densities.
At very low $T$$\lessim$50 K, these curves show a quadratic behaviour in T, consistent with the fact that   they   describe  a certain type of Fermi liquid- namely an extremely correlated Fermi liquid. This low T normal regime is  inaccessible  in most experiments since it overlaps with the superconducting phase.  In comparing with data we are therefore constrained  to  T$\gssim$50K,  without any  theoretical  limitation on the higher T side. Experimental data in most cases is  available in only a part of this potentially large extended region.

We note that LCCO, the electron doped \htsc system with $t'/t>0$, exhibits  a different  trend in the theoretical curves with enhanced regions showing $\rho(T)$$\sim$$T^2$ behaviour  compared to the hole doped  LSCO (and other compounds like BSLCO) where $t'/t<0$.
Theoretical curves for the  latter show   regions between 50K and $\sim$400K or a bit higher, where  $\rho(T)$ is linear with T, and then bend over  with an opposite curvature in a complicated crossover.  Much of the crossover  occurs for T$\gssim$350K, where  data is sparse.  In view of this situation  where data is available in a limited region of $T$,   inferring a  quasilinear behaviour $\rho(T)$$\sim$T from experiments, and theory  as well,  requires caution. At the minimum,  such a claim needs to be qualified in fairness by stating the range of T over  where it has been observed to  hold. Over a broader  range our almost T linear theoretical curves ultimately  bend over as seen in \figdisp{All-T},  providing a cautionary counterpoint.


\begin{figure}[H]
\begin{adjustwidth}{-1.5cm}{-1.5cm} 
\centering
{\includegraphics[width=.39\columnwidth]{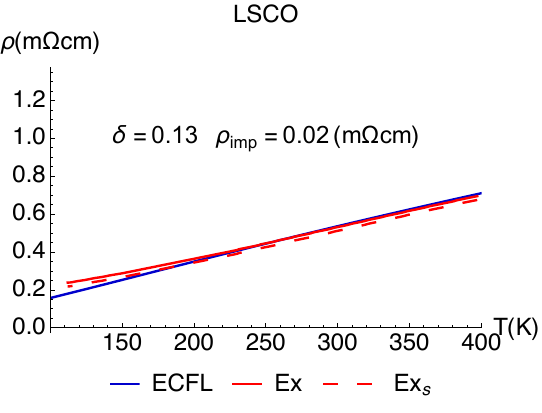}}
{\includegraphics[width=.39\columnwidth]{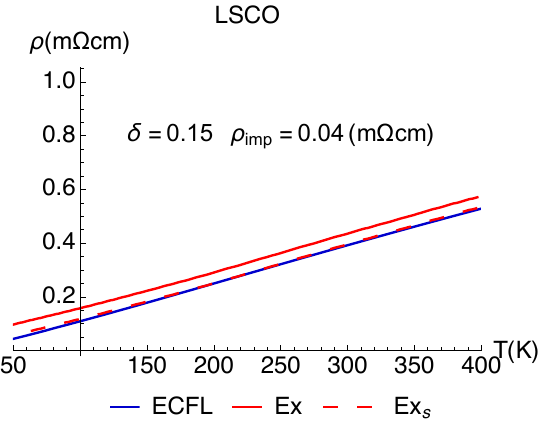}}
{\includegraphics[width=.4\columnwidth]{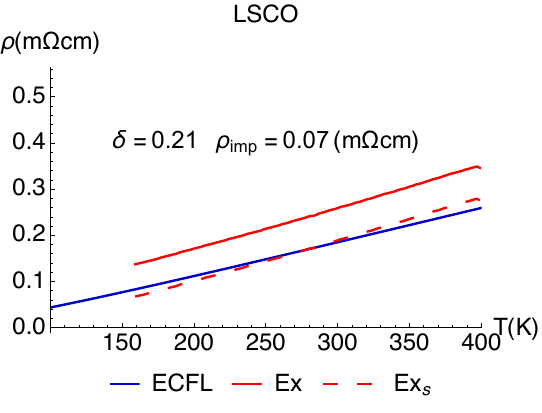}}
 \caption{\footnotesize    {\bf \S LSCO:} Theoretical resistivity for the hole doped  LSCO (t'/t$<$$0$) in blue, compared with the experiments in orange from Ref. \cite{Ando-1} at three  widely dispersed densities. The curve marked Ex is data obtained by digitizing the published figure,    and Ex$_s$  corrects for an estimated impurity resistance $\sim$2-5\% from the measured values. The data for $\delta$=0.21 at lower T was difficult to digitize due to overlapping curves in \cite{Ando-1}, but in the displayed range it appears to be linear  to the unaided eye. {\it Adapted from \cite{2dResults1},\copyright (2020) }
 \label{Fig-LSCO} }
 \end{adjustwidth}
 \end{figure}

\begin{figure}[tbhp]
\begin{adjustwidth}{-1.5cm}{-1.5cm} 
\centering
{\includegraphics[width=.39\columnwidth]{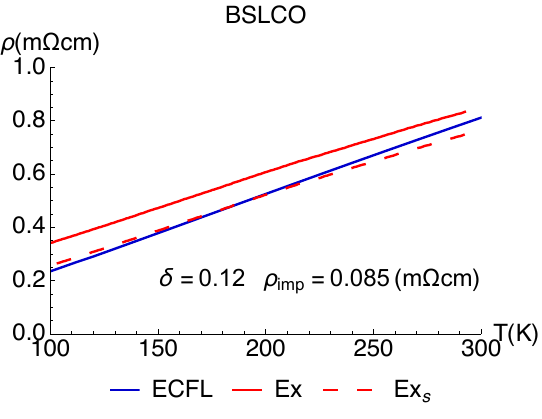}}
{\includegraphics[width=.39\columnwidth]{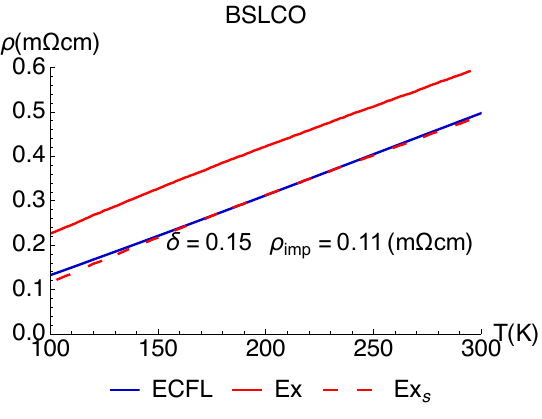}}
{\includegraphics[width=.39\columnwidth]{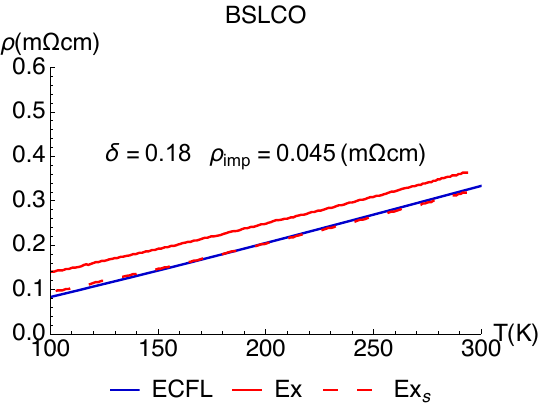}}
 \caption{\footnotesize {\bf \S BSLCO:}
 Theoretical resistivity for the hole doped   BSLCO (t'/t$<$$0$) in blue, compared with the experiments in orange from Ref. \cite{Ando-1} at denoted densities. The curve marked Ex is data obtained by digitizing the published figure,    and Ex$_s$  corrects for an estimated impurity resistance $\sim$10\% from the measured values.
 \label{Fig-BSLCO} {\it Adapted from \cite{2dResults1},\copyright (2020) }}
 \end{adjustwidth}
 \end{figure}

In \figdisp{Fig-LSCO} the data for LSCO \cite{Ando-1} at three densities $\delta=$0.13, 0.15 and 0.21 is compared with the theoretical  results. The single parameter used here   accounts for all 11  samples densities that are available in \cite{Ando-1} for this system. The three densities are in the underdoped, optimum doping and overdoped regimes and provide a representative set.
   In \figdisp{Fig-BSLCO}
 the data for BSLCO at densities $\delta$=0.12,0.15 and 0.18 from \cite{Ando-1} are compared with theoretical results. In both systems the density dependent range  of  variation of resistivity  is captured by the theory, which also accounts reasonably with the observed T dependence.

\begin{figure}[tbhp]
\begin{adjustwidth}{-1.5cm}{-1.5cm} 
\centering
{\includegraphics[width=.39\columnwidth]{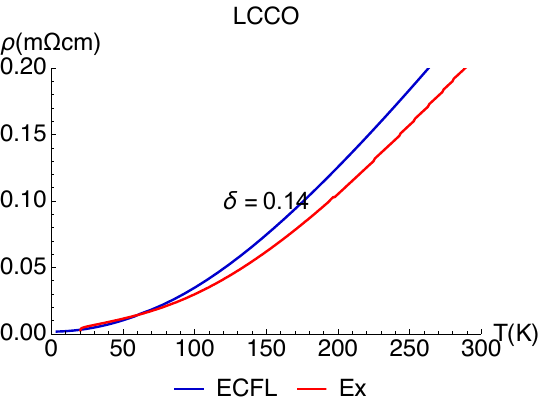}}
{\includegraphics[width=.39\columnwidth]{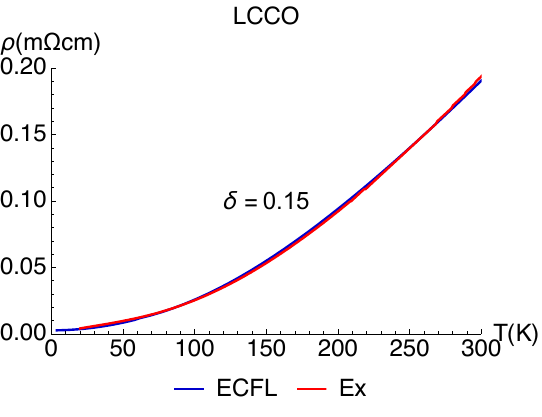}}
{\includegraphics[width=.39\columnwidth]{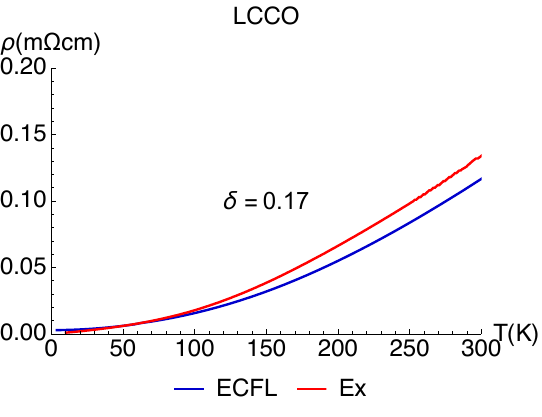}}
 \caption{\footnotesize {\bf \S LCCO:}  Theoretical resistivity for the electron doped  LCCO (t'/t$>$$0$) in blue, compared with the experiments in orange from Ref. \cite{LCCO} at denoted densities. The curve marked Ex is data corrected for impurity effects in \cite{LCCO}.  Note that the resistivity is overall of the type $\rho(T)\sim T^2$,  which is strikingly different from the behaviour seen in \figdisp{Fig-LSCO,Fig-BSLCO}.
 {\it Adapted from \cite{2dResults1},\copyright (2020) }
  \label{Fig-LCCO} }
  \end{adjustwidth}
 \end{figure}

In \figdisp{Fig-LCCO} the data for LCCO from \cite{LCCO} is compared with  theoretical results from ECFL using the  equations (\disp{eq27}-\disp{SR-2}). The equations are exactly the same as in \figdisp{Fig-LSCO,Fig-BSLCO}, but with a different set of parameters given in Table.~1. The  data with the impurity contribution already eliminated is   available in \cite{LCCO},  so it can be directly compared with theory. The three densities $\delta$=0.14,0.15 and 0.17 are closer together than the densities in \figdisp{Fig-LSCO,Fig-BSLCO}, but it is seen that the observed resistivity $\rho$$\sim$$T^2$ in all cases. The theoretical results have a  similar T dependence, and the quantitative agreement seems reasonable. It is worth emphasizing  that the same  ECFL theory equations (\disp{eq27}-\disp{SR-2}) employing  t'/t$>$0 used here, and  in \figdisp{All-T} ({\bf Right}), gives results that are quite different from those in \figdisp{Fig-LSCO,Fig-BSLCO} where we use t'/t$<$0.

\begin{figure}[tbhp]
\begin{adjustwidth}{-1.5cm}{-1.5cm} 
\centering
{\includegraphics[width=.45\columnwidth]{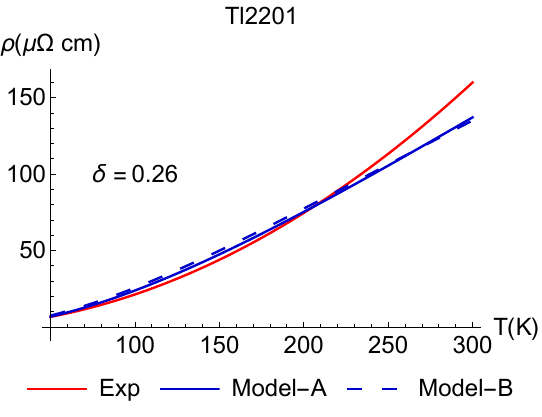}}
{\includegraphics[width=.45\columnwidth]{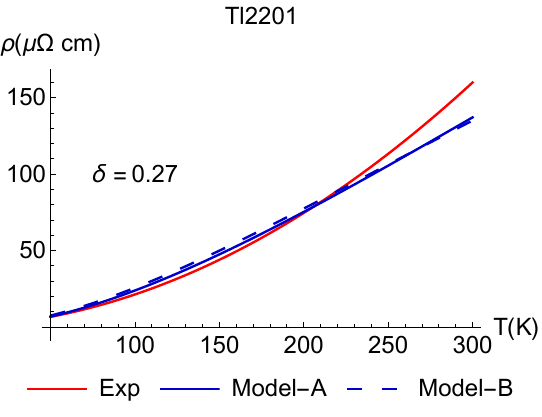}}
 \caption{\footnotesize {\bf \S Tl2201:} Theoretical resistivity for two possible set of hopping parameters that are quite distinct,   for  highly hole doped  Tl2201 (t'/t$<$$0$) in blue, compared with the experiments in orange from Ref. \cite{Cooper} at denoted densities. Model-A  uses t'/t=-043, t''/t=0.005 and t=1.82 eV, whereas Model-B uses t'/t=-0.237, t''/t=0.138 and t=1.053 eV, while producing the same shape of the Fermi surface. {\it Adapted from \cite{2dResults2},\copyright (2025) }
    \label{Fig-TICCO} }
    \end{adjustwidth}
 \end{figure}

In \figdisp{Fig-TICCO} we compare the data  on TlCCO from \cite{Cooper} with two sets of theoretical results from ECFL, which employ rather distinct values of the hopping parameters. This situation arises because the observed Fermi surface can be fitted equally well using two sets of band parameters given in Table.~1. This situation allowed us to test the hypothesis that the ECFL results should dependent  primarily  on the shape of the Fermi surface, and less so on the actual values of the hopping parameters. We see that both model parameters give   similar results and 
provide a fair account of the data up to $\sim$250K, beyond which theory is somewhat flatter  than the data.


\begin{figure}[H]
\centering
\begin{adjustwidth}{-1.5cm}{-1.5cm} 
{\includegraphics[width=.5\columnwidth]{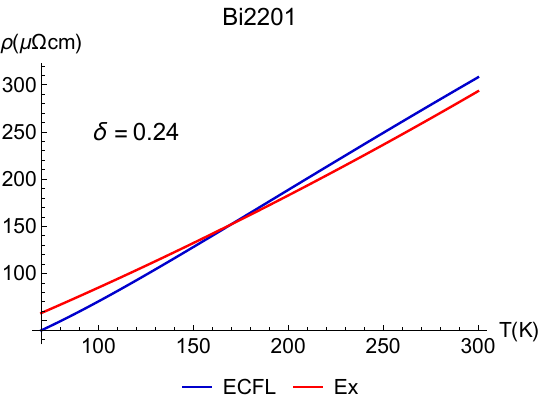}}
{\includegraphics[width=.5\columnwidth]{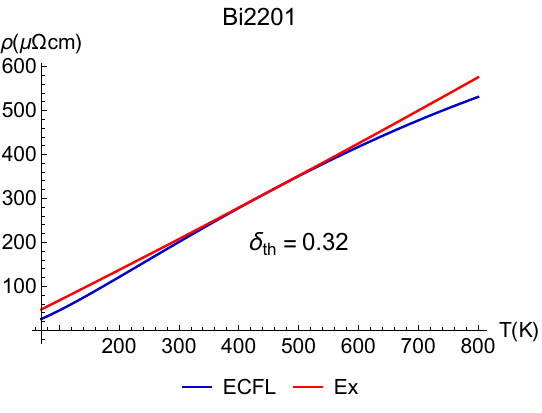}}
 \caption{\footnotesize {\bf \S Bi2201:}
 Theoretical resistivity   for  highly   hole  doped  Bi2201 (with t'/t$<$$0$) in blue, compared with the experiments in orange. {\bf [Left]}  Data is  from Ref. \cite{Hussey}. {\bf [Right]} Data is from Ref.~\cite{Martin,Fiory}. Note the   wide  range of T quoted in this early data. The quoted T$_c$ of this sample  \cite{Martin} translates to  an estimated $\delta$$\sim$0.259, if one  uses a phenomenological  T$_c$-$\delta$ relation. That estimate 
 differs slightly from the density found later \cite{Hussey} in  samples with similar T$_c$. In this plot we  assumed  a slightly greater $\delta_{th}$=0.32, which gives a better theoretical fit \cite{Fiory-Thanks}.  {\it Adapted from \cite{2dResults2},\copyright (2025) }
 \label{Fig-Bi2201} }
 \end{adjustwidth}
 \end{figure}

In \figdisp{Fig-Bi2201}({\bf Left}) we compare the data for highly hole doped Bi2201 from \cite{Hussey} with the theoretical results. In \figdisp{Fig-Bi2201}({\bf Right}), we compare  the early data from \cite{Martin,Fiory} mentioned in the Introduction,  with theoretical results assuming that the sample density is $\delta$=0.32. The authors of \cite{Martin,Fiory} do not quote a density, but 
 an approximate value $\delta$=0.259  is suggested by phenomenology relating the superconducting T$_c$ to density. Our assumed value  amounts to a guess for the  magnitude of density  such that the resistivity fits  reasonably well to theory. Our choice also  seems to be consistent with the recent work of \cite{Hussey}, who quote resistivity of samples  with similar magnitudes, in a similar range of density to our   guess.

\section{Summary \label{Summary}}

We next  comment briefly  on  goal (d) relating to  superconductivity. It must be kept in view that
 there is currently no rigorous proof of high temperature superconductivity  arising from  the \tJ model, without the aid of extra terms beyond those in \disp{HtJ}. Therefore, and despite its popularity  inspired by initial efforts \cite{Hirsch,Anderson,BZA,Kotliar,Gros}, this goal comes with some  intrinsic uncertainty.   In \cite{ECFL-SC}, we have adapted the ECFL framework to investigate the possibility of superconductivity within the \tJ model by allowing for superconducting order, generalizing the approach  of Go`rkov\cite{Gorkov,AGD,Engelsberg}.  
A generalized criterion for determining T$_c$ is formulated which involves the Greens function of the correlated electrons. Since the Cooper pairing   involve  quasiparticles, the small values of  the quasiparticle weight Z found in this theory inhibit   pairing. This is seen most clearly  by comparing with the toy model consisting  of an uncorrelated model with the same interaction\cite{ECFL-SC}, where one finds much higher T$_c$$\sim$$10^4$K. The  results reported in \cite{ECFL-SC}  are  encouraging. Calculations  yield a dome-type superconducting phase with d-wave order, with T$_c$$\sim$$10^2$K. However   further work is needed to clarify questions regarding  the range of parameters that support superconductivity,  which is currently somewhat restrictive.

Regarding the main  question (Q1) in Section~\ref{Intro}, we see that the ARPES line shapes are not as mysterious as they appeared at first. The ECFL theory gives a line shapes that are  non-Lorentzian,  which closely resemble available  data. The theory also gives a simple and testable  prediction about kinks, and the temperature dependence of spectral peaks,  which  are amenable to experimental verification.

In our discussion   we have outlined the framework of the ECFL theory.   The comparisons with data  presented above suggests that the  goals (a-c) listed in the introduction are close to completion. The results on transport in Section.~\ref{Transport} give a fair   account of the  T and density dependence of the  resistivity observed in the full set of single layer High$T_c$ compounds. More broadly, the data sets analyzed in this work  argue against a universal T-linear resistivity characterizing all High T$_c$ materials.  A quasi-linear resistivity is indeed seen, but  only  over a finite range of T, which  in turn is 
density dependent. These  results are seen to be  consistent with the  description provided by the ECFL calculations. 

An outgrowth of (Q1) is the  question of   whether one can make a determination of the nature of the quantum state in the normal phase of High T$_c$ systems, on the basis of the resistivity and its T dependence  \cite{LCCO}. To specifically address this question, it seems desirable  to throw the net somewhat wider. More specifically, it would be most helpful to  acquire information on the  imaginary part of the single particle self energy $\Sigma(\vec{k}_F,\omega)$, at the lowest $\omega$.   At the very lowest $\omega$, Landau Fermi liquids and    ECFL calculations  show in common, an $\omega^2$ behavior of $\Im \Sigma(\vec{k},\omega)$,  while exhibiting substantial differences detailed  elsewhere. Other, more radically different quantum states, are expected to show different powers of  $\omega$.  
  Motivated by this  question,  a recent work \cite{Shastry-Selfenergy} 
 proposes a new approach for reconstructing the $\Im \, \Sigma(\vec{k},\omega)$  from the  spectral function $A(\vec{k},\omega)$, obtainable in principle from photoemission experiments. This uses the formula
\beq
- \frac{1}{\pi} \Im \, \Sigma(\vec{k},\omega) = \frac{A(\vec{k},\omega)}{\{\pi A(\vec{k},\omega) \}^2+ \Phi_k^2(\omega)}
\eeq
where $\Phi_k(\omega)$  is the Hilbert transform of $A(\vec{k},\omega)$.
A simple theorem shows further  that errors in this  process,  arising mainly  from determining $\Phi_k$ from $A(\vec{k},\omega)$  measured over a finite and usually small range of $\omega$, are not  fatal to the inversion. Several favourable factors are at play-  in certain well defined situations these can make the errors   {\em vanishingly small} as $\omega$$\to$0, i.e. in  the   region most relevant for answering the original question.

In characterizing the results of the ECFL theory,
we have noted the emergence of several low temperature scales in the calculated results. In order to understand their origin at a  qualitative level, we observe that the Dyson self energy is a  composite object in the theory (see the discussion after \disp{Self-Psi,Self-Phi}).  The T dependence of $\bmu$ and $u_0$ entering this construction further  adds to the complexity,   resulting  in the  generation of    multiple low T scales.
 These scales are  visible across different observables- including the resistivity crossover  and  the high entropy release relative to an uncorrelated band at low T, indicating a reduced effective Fermi energy\cite{Wenxin-Paper}(Figs~1,12)-  as well as  the sensitive T dependences seen in   \figdisp{Peaks-1,Peaks-2,RHall,EDC-1,EDC-2}.
These   dominate the resistivity results displayed in Sec(\ref{Resistivity}), and are also seen in Hall transport, thermal,  and  spectral properties.  Another  characteristic is  the prominent reduction of the quasiparticle weight $Z\ll1$, which is  controlled  by  the density and the hopping parameters of the model \figdisp{Zdelta}. 

In both characteristics,  we see that changing the hopping parameters leads to substantial changes in the theoretical  results. Since the hopping  parameters are material-specific, even within a given class,  these add up to give an unprecedented  situation with    strongly material-specific  effect of correlations.
These characteristizations   define  the {\em Extremely Correlated Fermi Liquid} (ECFL) {\em state}.

 Taken together, the    work reported here  suggests  that the ECFL  theory provides a consistent and quantitative  description of the observed transport and spectral properties of the important class of single-layer High $T_c$ materials.

\section{ Acknowledgements}

I am  grateful to numerous colleagues for helpful discussions and comments, which greatly benefitted this work \footnote{ Based on a Colloquium  at  the Max Planck Institute for the Physics of Complex Systems, Dresden,  by the author  on 10th November 2025.}. I particularly thank several collaborators who have contributed directly to the work reported here. They include, chronologically, (i) students and postdocs Daniel Hansen, Edward Perepelitsky, Peizhi Mai, Kazue Matsuyama,  Wenxin Ding, Michael Arciniaga and Samantha Shears; and, alphabetically, (ii) colleagues Antoine Georges, Gey-Hong Gweon, Alex Hewson, Ehsan Khatami, H. R. Krishnamurthy, Jernej Mravlje, Marcos Rigol, Steve White, and Rok \v{Z}itko.

I acknowledge enlightening discussions 
 with Professor Phillip Anderson  in 2010-11, on the need for a   fresh theoretical   approach to the theory of strong correlations, and  particularly on the contents of  Sec 2.1.



\begin{thebibliography}{99}
\bibitem{ECFL} B. S. Shastry, Extremely Correlated Fermi Liquids ,   Phys. Rev. Letts. {\bf 107},  056403 (2011).


\bibitem{tJ-Hubbard-Connection}
The  \tJ  model can be obtained from the large U Hubbard model, by a canonical transformation that eliminates doubly occupied sites.  It involves  the  neglect of  a  three body term  which becomes very small near half filling.
 A.B.Harris, R.V.Lange,Phys.Rev.{\bf 157}, 295 (1967); K.A.Chao, J.Spalek, A.M. Oles, J.Phys.{\bf C10}, L271(1977).

\bibitem{ECFLall} Compilation of reprints on the Extremely Correlated Fermi Liquid  theory (with comments): \url{https://escholarship.org/uc/item/9pf2t069 }



\bibitem{Martin}S. Martin, A. T. Fiory, R. M. Fleming, L. F. Schneemeyer, and J. V. Waszczak, { Normal-state transport properties of $Bi_{2+x}Sr_{2-y}CuO_{6 \pm \delta}$ crystals},  Phys. Rev. B {\bf 41}, 846 (1990).


\bibitem{Arko} B. G. Wells, Z. X. Shen, D. S. Dessau, W. E. Spicer, C. G. Olson, D. B. Mitzi, A. Kaputalnik, R. S. List and A. Arko, Angle-Resolved Photoemission Study of Bi$_2$Sr$_2$CaCu$_2$O$_{8 +\delta}$: Metallicity of the Bi-O Plane, Phys. Rev. Letts. {\bf 65}, 3056 (1990).

\bibitem{Landau-Pomeranchuk} L. D. Landau, I. Y. Pomeranchuk, { On the Properties of Metals at Very Low Temperatures},    Sov. J. Exp. Theor. Phys.  {\bf 7}, 379, (1937).

\bibitem{Wilkins}  W. E. Lawrence and J. W. Wilkins, Electron-electron scattering
in the transport coefficients of simple metals, Phys. Rev. B {\bf 7},
2317 (1973).

\bibitem{MJRice} M. J. Rice, { Electron-Electron Scattering in Transition Metals},  Phys. Rev. Letts. {\bf 20}, 1439 (1968).

\bibitem{Miyake-Varma}  K. Miyake, T. Matsuura and C. M. Varma, { Relation Between Resistivity and Effective Mass in Heavy-Fermion and A-15 Compunds}, Sol. State. Comm. {\bf 71}, 1149 (1989).





\bibitem{Luttinger-Ward}  J.M. Luttinger and  J.C. Ward, Ground-State Energy of a Many-Fermion System. II,  Phys. Rev. 118 (1960) 1417.

\bibitem{DMFT} A. Georges, G. Kotliar, W. Krauth, and M. J. Rozenberg,
Dynamical mean-field theory of strongly correlated fermion
systems and the limit of infinite dimensions, Rev. Mod. Phys.
68, 13 (1996).
\bibitem{NRG} K. G. Wilson, The renormalization group: Critical phenomena
and the Kondo problem, Rev. Mod. Phys. 47, 773 (1975).

\bibitem{NRG2} H. R. Krishnamurthy, J. W. Wilkins, and K. G. Wilson,
Renormalization-group approach to the Anderson model of
dilute magnetic alloys. I. Static properties for the symmetric
case, Phys. Rev. B 21, 1003 (1980); 
Renormalization-group approach to the Anderson model of
dilute magnetic alloys. II. Static properties for the asymmetric
case, Phys. Rev. B 21, 1044 (1980).

\bibitem{fn1}This   theorem \cite{Luttinger-Ward} is based on the assumption that  perturbation theory to all orders is  convergent. It asserts that under that assumption,  the  volume (or area in 2-d) of the Fermi surface at $T=0$ is unchanged from that of the Fermi gas with the same band structure parameters. There are  further  subtleties relating to finite T definitions of the Fermi surface, and indeed of how one defines the Fermi surface at all \cite{Shastry-Luttinger,Oshikawa-Luttinger}.

\bibitem{Shastry-Luttinger}  B.S. Shastry, Fermi Surface Volume of Interacting Systems,  arXiv:1808.00405v3, Annals of  Physics {\bf 405}, 155 (2019). 
https: //doi.org/ 10.1016/j.aop.2019.03.016


\bibitem{Oshikawa-Luttinger}   M. Oshikawa, Topological approach to Luttinger’s theorem and the Fermi surface of a Kondo lattice,  Phys. Rev. Lett. {\bf 84},  3370 (2000).

\bibitem{Izumov} Yu. A. Izyumov, B. M. Letfulov, J. Phys.: Condens. Matter 2 (1990) 8905–8923.

\bibitem{ECQL}  B. S. Shastry, Extremely Correlated Quantum Liquids,  Phys. Rev. B {\bf 81}, 045121 (2010);  arXiv.org:0911.4327.



\bibitem{PathIntegral} B. S. Shastry, Theory of extreme correlations using canonical Fermions and path integrals, arXiv:1312.1892 (2013),  Ann. Phys. {\bf 343}, 164-199 (2014).  DOI:http://dx.doi.org/10.1016/j.aop.2014.02.005.  (Erratum) Ann. Phys. Vol.  373, 717-718 (2016).  \\ DOI:http://dx.doi.org/10.1016/j.aop.2016.08.015.

\bibitem{PauliPrinciple} D. J. Amit and H. Keiter, Functional Integral Approach to the Magnetic Impurity Problem: The Superiority of the
Two-Variable Method, Jour. Low. Temp. Phys. {\bf 11}, 603 (1973). 


\bibitem{ECFL-DMFT}   R. \v{Z}itko, D. Hansen, E. Perepelitsky, J. Mravlje, A. Georges and B. S. Shastry, Extremely correlated Fermi liquid theory meets Dynamical mean-field theory:
Analytical insights into the doping-driven Mott transition, arXiv:1309.5284 (2013), Phys. Rev. {\bf B 88}, 235132 (2013).


\bibitem{2dResults1}  B. S. Shastry and P. Mai, Aspects of the Normal State Resistivity of Cuprate Superconductors, arXiv:1911.09119; Phys. Rev. B{\bf 101},115121(2020); DOI:  https://doi.org/10.1103/PhysRevB.101.115121

\bibitem{2dResults2}  S. Shears, M. Arciniaga and B. S. Shastry, Aspects of the normal state resistivity of cuprate superconductors $Bi_2Sr_2CuO_{6+x}$, $Tl_2Ba_2CuO_{6+x}$ and $HgBa_2CuO_{4+x}$, Phys. Rev. B {\bf 111}, 245146 (2025); https://DOI: 10.1103/89cj-5qhs;  arXiv:2502.00293.
\bibitem{ECFL-Edward-Shastry}  E. Perepelitsky and B. S. Shastry, Diagrammatic $\lambda$ series for extremely correlated Fermi liquids, arXiv: 1410.5174, Ann. Phys. {\bf 357}, 1 (2015). doi:  https://doi.org/10.1016/j.aop.2015.03.010

\bibitem{Comments-Equations}  These equations can be found in \cite{2dResultsFirst,Mike}. These correspond to the ``minimal theory'' of \cite{ECFL-Edward-Shastry} Eqs. (63-67). In earlier work, e.g. \cite{ECFL}, slightly different equations are reported. The reported equations added extra terms to the equations of motion, which vanish in an exact treatment but not in the approximate version. The motivation was to obtain great symmetry in the equations.  In \cite{ECFL-Edward-Shastry,2dResultsFirst,2dResults1,2dResults2,Mike} we dropped  these added terms for simplicity.

\bibitem{Mike}
Code of the  calculations of spectral functions and resistivity in 
\cite{2dResults1,2dResults2} is available in two formats:\\
S. Shears, M. Arciniaga and B. S. Shastry, 
Aspects of the normal state resistivity of cuprate superconductors Bi2201, Tl2201 and Hg1201 \\
  Zenodo:  \url{ https://doi.org/10.5281/zenodo.18085923} \\
  Github: \url{ https://github.com/marcinia-sudo/2DECFL2ndOrder}

\bibitem{Sam}   Calculated spectral functions, electron self-energy and resistivity used in \cite{2dResults1,2dResults2} is available in two formats:\\
S. Shears, M. Arciniaga and B. Sriram Shastry, 
Aspects of the normal state resistivity of cuprate superconductors Bi2201, Tl2201 and Hg1201,\\
Zenodo:  \url{ https://doi.org/10.5281/zenodo.18350589} \\
Github: \url{https://github.com/s-shears/}

\bibitem{ECFL-AIM} ``Extremely Correlated Fermi Liquid study of the $U=\infty$ Anderson Impurity Model'', B. S. Shastry, E. Perepelitsky and A. C. Hewson, arXiv:1307.3492 [cond-mat.str-el], Phys. Rev. {\bf B 88}, 205108 (2013).

\bibitem{Zlatic} B. Horvati\'c and V. Zlati\' c, Perturbation expansion for the asymmetric Anderson Hamiltonian II. General asymmetry, Phys.
Status Solidi 111, 65 (1982)

\bibitem{Zitko} Rok \v{Z}itko, H. R. Krishnamurthy and B. S. Shastry, Reversal of particle-hole scattering-rate asymmetry in Anderson impurity model,  arXiv:1807.11343, Phys. Rev.  B{\bf  98}, 161121(R) (2018). DOI:  https://doi.org/10.1103/PhysRevB.98.161121


\bibitem{Sign-Thermopower}K. Haule and G. Kotliar, Thermoelectrics near the Mott
localization-delocalization transition, in Properties and Appli-
cations of Thermoelectric Materials, edited by V. Zlati\'c and A.
C. Hewson (Springer, Dordrecht, 2009), p. 119.

\bibitem{DMRG} S. R. White, Early Days of DMRG, Nat. Phys. Rev. {\bf 5}, 264 (2023).



\bibitem{ECFL-1d}  P. Mai, S. R. White and B. S. Shastry,   The t-t'-J model in one dimension using extremely correlated Fermi liquid theory and time dependent density matrix renormalization group, arXiv:1712.05396, Phys. Rev. B98, 035108 (2018). DOI: 10.1103/PhysRevB.98.035108.

\bibitem{ECFL-InfiniteD} E. Perepelitsky and B. S. Shastry, ECFL in the limit of infinite dimensions,  arXiv: 1309.5373 (2013), Annals of Physics {\bf 338}, 283-301  (2013).

\bibitem{2dResultsFirst}   B. S. Shastry and P. Mai, Extremely Correlated Fermi Liquid theory of the $t$-$J$ model in 2 dimensions: Low energy properties,arXiv:1703.08142, New Jour. Phys. {\bf 20} 013027 (2018). DOI: https://doi.org/10.1088/1367-2630/aa9b74

\bibitem{Mai-Shastry}
P. Mai and B. S. Shastry, Extremely correlated fermi liquid of $t$-$J$ model in two dimensions,  arXiv:1808.09788;  Phys. Rev. B{\bf 98}, 205106 (2018). DOI:  https://doi.org/10.1103/PhysRevB.98.205106.


\bibitem{Potthoff} M. Potthoff, Non-perturbative construction of the Luttinger-Ward functional,  arXiv.cond-mat/0406671; Condens. Matter Phys. 9, 557 (2006).

\bibitem{deDominicis} C. de Dominicis and P.C. Martin, Stationary Entropy Principle and Renormalization in Normal and Superfluid Systems. I. Algebraic Formulation, J. Math. Phys. {\bf 5} 14 (1964).

\bibitem{Gweon} G.-H. Gweon, B. S. Shastry and G. D. Gu,  Extremely Correlated Fermi Liquid Description of Normal State ARPES in Cuprates, arXiv:1104.2631 (2011), Phys. Rev. Letts. 107, 056404 (2011). DOI: 10.1103/PhysRevLett.107.056404


\bibitem{casey}  P. A. Casey, J. D. Koralek, N. C. Plumb, D. S. Dessau and P. W. Anderson,  Accurate theoretical fits to laser-excited photoemission spectra in the normal phase of high-temperature superconductors  et al, Nature Physics 4, 210 (2008).

\bibitem{Kazue} K. Matsuyama, E. Perepelitsky and B. S. Shastry,     Origin of kinks in the energy dispersion of strongly correlated matter, Phys. Rev. B {\bf 95}, 165435 (2017)
\bibitem{Kaminsky} A. Kaminski, M. Randeria, J. C. Campuzano, M. R. Norman,
H. Fretwell, J. Mesot, T. Sato, T. Takahashi, and K. Kadowaki,
Phys. Rev. Lett. 86, 1070 (2001).

\bibitem{voruganti}
  H. Fukuyama, H. Ebisawa, and Y. Wada: Prog. Theor. Phys. {\bf 42} 494 (1969); H. Kohno and K. Yamada, Prog. Theor. Phys. {\bf 80} 623 (1988);P. Voruganti, A. Golubentsev and S. John, Phys. Rev. {\bf B 45}, 13945 (1992).



\bibitem{Tremblay} L-F Arsenault and A.M. S. Tremblay
Phys. Rev. {\bf B 88}, 205109 (2013)




\bibitem{Takagi} H Takagi, T Ido , S Ishibashi, M Uota, S Uchida  and Y Tokura  Phys. Rev. B 40, 2254 (1989).

\bibitem{Ong}  T R Chien, Z Z Wang  and N P  Ong  Phys. Rev. Lett. 67, 2088 (1991); 
N P Ong  and P W  Anderson   Phys. Rev. Lett. 78, 977 (1997).

\bibitem{Shastry-Hall} B. S. Shastry, Electrothermal transport coefficients at finite frequencies,  Rep. Prog. Phys. {\bf 72}, 
016501 (2009).

\bibitem{Samantha-Hall} S. Shears, B. S. Shastry and M. Arciniaga, Extremely Correlated Fermi Liquid Theory - $\cot$ $\Theta_H$ and Related Results, SCFS25 Conference on Current Trends in Strongly Correlated and Frustrated Systems, 10-14 November 2025, MPIKS, Dresden, Germany; and in preparation. 

\bibitem{Names}We will refer to the different families by their popular  acronyms
 LSCO (La$_{2-x}$Sr$_x$CuO$_4$), 
BSLCO (Bi$_2$Sr$_{2-x}$La$_x$CuO$_{6+\delta}$),
 LCCO (La$_{2-x}$Ce$_x$CuO$_4$),
  Bi2201 (Bi$_2$Sr$_2$CuO$_{6+x}$),
   Tl2201 (Tl$_2$Ba$_2$CuO$_{6+x}$),
    Hg1201 (HgBa$_2$CuO$_{4+x}$).

\bibitem{LCCO}  T. Sarkar, R. L. Green and S. Das Sarma,{ Anomalous normal-state resistivity in superconducting $La_{2-x} Ce_x CuO_4$: Fermi liquid or strange metal?}, Phys. Rev. B {\bf 98}, 224503 (2018).

\bibitem{Cooper} J. R. Cooper, J. C. Baglo, C. Putzke and A. Carrington, { Thermoelectric power of overdoped Tl2201 crystals: charge density waves and T$^1$ and T$^2$ resistivities },  Supercond. Sci. Technol. {\bf 37}, 015017 (2024).

\bibitem{Hussey} M. Berben, S. Smit, C. Duffy , Y.-T. Hsu , L. Bawden, F. Heringa, F. Gerritsen , S. Cassanelli, X. Feng, S. Bron, E. van Heumen, Y. Huang, F. Bertran , T. K. Kim , C. Cacho,
A. Carrington, M. S. Golden, and N. E. Hussey, { Superconducting dome and pseudogap endpoint in Bi2201},
 Phys. Rev. Materials {\bf 6}, 044804 (2022).

\bibitem{Ando-1}Y. Ando, Y. Kurita, S. Komiya, S. Ono, and K. Segawa, {Electronic Phase Diagram of High-Tc Cuprates from a Mapping of the In-Plane Resistivity Curvature}, Phys. Rev. Lett. {\bf 93}, 267001 (2004).

\bibitem{Fiory} A. T Fiory, S. Martin, R. M. Fleming, L. F. Schneemeyer, J. V. Waszczak, A. F. Hebard and S. A. Sunshine,{ Transport, Tunneling, X-Ray, and penetration depth studies of superconducting $Bi_{2+x}Sr_{2-y}CuO_{6\pm\delta}$ crystals},  Physica {\bf C}, 162-164, 1195 (1989).

\bibitem{Cooper2} J. R. Cooper, J. C. Baglo, C. Putzke and A. Carrington, { Thermoelectric power of overdoped Tl2201 crystals: charge density waves and T$^1$ and T$^2$ resistivities },  Supercond. Sci. Technol. {\bf 37}, 015017 (2024).


\bibitem{Yamamoto} A. Yamamoto, W. Hu and S. Tajima, { Thermoelectric power and resistivity of $HgBa_2CuO_{4+\delta}$ over a wide doping range}, 
Phys. Rev. B {\bf 63}, 024504 (2000).


\bibitem{NCCO} P. K. Mang, S. Larochelle, A. Mehta, O. P. Vajk, A. S. Erickson, L. Lu, W. J. L. Buyers, A. F. Marshall, K. Prokes, and M. Greven, {\em Phase decomposition and chemical inhomogeneity in $Nd_{2-x}Ce_xCuO_{4\pm \delta}$}, Phys. Rev. B {\bf 70}, 094507 (2004).

\bibitem{Fiory-Thanks}  The labels of the theoretical and experimental curves was inadvertently interchanged in the related  Fig.~4 of  \cite{2dResults2}. I thank Dr A. T. Fiory for  pointing this  out this error.  










\bibitem{Hirsch}J. E. Hirsch, Attractive Interaction in Pairing in Fermion Systems with Strong On-Site Repulsion,  Phys. Rev. Letts. {\bf 54}, 1317 ( 1985).
\bibitem{Anderson} P. W. Anderson, The Resonating Valence Bond State in La2CuO4 and Superconductivity,  Science {\bf 235}, 1196 (1987).
\bibitem{BZA} G. Baskaran, Z. Zou and P. W. Anderson,  The Resonating Valence Bond State and High T$_c$ Superconductivity- A Mean Fied Theory,  Sol. St. Comm. {\bf 63}, 973 (1987).
\bibitem{Kotliar}  G.Kotliar, Resonating valence bonds and d-wave superconductivity, Phys.Rev. B {\bf 37} 3664 (1988).
\bibitem{Gros} C. Gros, Physics of projected wavefunctions,  Ann. Phys. {\bf 189}, 53 (1989).

\bibitem{ECFL-SC}   B. S. Shastry,  Extremely Correlated Superconductors, arXiv:2102.08395; Annals of Physics, {\bf 434}, 168614 (2021); \\ https://doi.org/10.1016/j.aop.2021.168614
\bibitem{Gorkov} L. P. Gor'kov, On the energy spectrum of Superconductors, Sov. Phys. JETP {\bf  7}, 505 (1958).
\bibitem{AGD} A. A. Abrikosov, L.  Gor'kov and I. Dzyaloshinski, {\em Methods of Quantum Field Theory in Statistical Physics },  Prentice-Hall,
Englewood Cliffs, NJ (1963).

\bibitem{Engelsberg} S. Engelsberg, Functional Derivative Techniques in the Theory of Superconductivity,  Phys. Rev. {\bf 126}, 1251 (1962).

\bibitem{Shastry-Selfenergy} B. S. Shastry, Method for reconstructing the self-energy from the spectral function, Phys. Rev. B {\bf 110} 155149 (2024).

\bibitem{Wenxin-Paper}  W. Ding, R. \v{Z}itko, P. Mai, E. Perepelitsky and B. S. Shastry,  Strange metal from Gutzwiller correlations in infinite dimensions, Phys. Rev. B {\bf 96}, 054114 (2017).

\end{thebibliography}
 \end{document}